\documentclass[12pt]{article}

\usepackage{graphicx}
\usepackage{amsmath}
\usepackage{amssymb}
\usepackage{amsfonts}
\usepackage{graphicx}
\usepackage{mathrsfs}

\newcommand{\be}{\begin{equation}}
\newcommand{\ee}{\end{equation}}
\newcommand{\bea}{\begin{eqnarray}}
\newcommand{\eea}{\end{eqnarray}}
\newcommand{\nnb}{\nonumber}

\newcommand{\tr}{\mathrm{tr}}

\newcommand{\pd}[2]{\frac{\partial #1}{\partial #2}}

\newcommand{\om}{\omega}
\newcommand{\Om}{\Omega}
\newcommand{\re}{\mathrm{Re}}
\newcommand{\im}{\mathrm{Im}}	
\newcommand{\cl}[1]{\mathcal{#1}}

\newcommand{\sst}{SU(3)$\times$SU(3) }

\def\slas#1{\rlap{\begin{picture}(10,10)(-5,0)
\put(0,0){\line(2,1){15}}
\end{picture}} #1 }

\def\slashH{\rlap{\begin{picture}(10,10)(0,1)
\put(0,0){\line(1,1){10}}
\end{picture}}H }

\def\slashD{\rlap{\begin{picture}(10,10)(0,1)
\put(0,0){\line(1,1){10}}
\end{picture}}D }

\def\slashPhipm{\rlap{\begin{picture}(10,10)(0,1)
\put(0,0){\line(1,1){10}}
\end{picture}} \Phi^0_\pm }

\def\slashPhiPlus{\rlap{\begin{picture}(10,10)(0,1)
\put(0,0){\line(1,1){10}}
\end{picture}} \Phi^0_+ }

\def\slashPhiMinus{\rlap{\begin{picture}(10,10)(0,1)
\put(0,0){\line(1,1){10}}
\end{picture}} \Phi^0_- }

\catcode`\@=11
\def\marginnote#1{}

\def\titlepage{\@restonecolfalse\if@twocolumn\@restonecoltrue\onecolumn
     \else \newpage \fi \thispagestyle{empty}\c@page\z@
        \def\thefootnote{\fnsymbol{footnote}} }

\def\endtitlepage{\if@restonecol\twocolumn \else  \fi
        \def\thefootnote{\arabic{footnote}} \setcounter{footnote}{0}}

\catcode`@=12 \relax

\makeindex

\begin{document}
\topmargin -1.4cm
\begin{titlepage}
\begin{flushright}
LPTENS-08/20\\
ROM2F/2008/05\\
April 2008
\end{flushright}
\vskip 2.5cm
\begin{center}
{\Large\bf Reducing democratic type II supergravity}\\
\vspace{3mm} {\Large \bf on SU(3)$\times$SU(3) structures}
\vskip 1.3cm 
{\bf Davide Cassani}\\
\vskip.6cm 
Laboratoire de Physique Th\'eorique,
\'Ecole Normale Sup\'erieure - CNRS UMR8549\footnote{
\hbox{unit\'e mixte du CNRS et de l'\'Ecole Normale Sup\'erieure 
associ\'ee \`a l'UPMC Univ. Paris 06} }
\\
24 rue Lhomond, 75231 Paris Cedex 05, France\\
\vskip .3cm
Dipartimento di Fisica, Universit\`a di Roma ``Tor Vergata''\\
Via della Ricerca Scientifica, 00133 Roma, Italy\\
\vskip .3cm 
{\tt davide.cassani@lpt.ens.fr}
\end{center}

\vskip 1.2cm

\begin{center}
{\bf Abstract}
\end{center}
\begin{quote}

Type II supergravity on backgrounds admitting SU(3)$\times$SU(3) structure and general fluxes is considered. Using the generalized geometry formalism, we study dimensional reductions leading to $N=2$ gauged supergravity in four dimensions, possibly with tensor multiplets. In particular, a geometric formula for the full $N=2$ scalar potential is given. Then we implement a truncation ansatz, and derive the complete $N=2$ bosonic action. While the NSNS contribution is obtained via a direct dimensional reduction, the contribution of the RR sector is computed starting from the democratic formulation and demanding consistency with the reduced equations of motion.
\end{quote}
\end{titlepage}
\setcounter{footnote}{0} \setcounter{page}{0}
\newpage
%

\section{Introduction}

\setcounter{equation}{0}

Dimensional reductions of type II theories can lead to $N=2$ supergravities in four dimensions. The basic well-known realization consists of compactifications on Calabi-Yau three-folds with no fluxes, in which case the $N=2$ effective action is ungauged, and contains hyper- and vector-multiplets, in addition to the gravitational one \cite{BodnerCadavidFerrara, IIBonCY}. The introduction of NS and RR fluxes in the higher dimensional background is described by a deformation of this four dimensional theory in which some specific isometries of the hyperscalar quaternionic manifold are gauged\footnote{A thorough account on gauged and ungauged 4d $N=2$ supergravity can be found in$\:$\cite{N=2review}. Refs.$\:$\cite{FluxReviews} are recent reviews on flux compactifications.} \cite{PolchinskiStrominger, Michelson, DallAgataHfluxes, LouisMicu, KachruKashaniPoor, TaylorVafa}. A consistent formulation in the presence of a complete set of RR fluxes requires the introduction of massive tensor multiplets \cite{LouisMicu}. 

Four dimensional theories with more complex gaugings can be derived extending the class of internal geometries beyond the Calabi-Yau domain. In recent years considerable efforts have been directed to the study of compactifications on manifolds with SU(3) structure (with restriction to $N=2$ reductions of type II, see \cite{GurLouisMicuWaldr, GurrMicuIIBhalfFlat, GaugingHeisenberg, TomasMirrorSymFl, GLW1, ChuangKachruTomasiello, HousePalti, MinasianKashani, KashaniPoorNearlyKahler}). This class of manifolds shares with the Calabi-Yau the existence of a globally defined and nowhere vanishing spinor, but is more general since such spinor needs not being covariantly constant in the Levi-Civita connection. A further motivation to study SU(3) structure manifolds is that they arise as mirror-symmetric duals of Calabi-Yau backgrounds with NS fluxes \cite{GurLouisMicuWaldr}. 

However, manifolds with strictly SU(3) structure are not the only candidates potentially leading to $N=2$ in four dimensions. Indeed, if a globally defined internal spinor $\eta$ is clearly needed in order to decompose the two type II susy parameters under Spin(9,1)$\,\to\,$Spin(3,1)$\times$ Spin(6) and preserve eight supercharges in 4d, there is also the possibility to employ {\it a pair} of internal spinors $\eta^1$ and $\eta^2$ in this decomposition: one for each of the ten dimensional susy parameters. The topological requirement associated with this situation is then that the internal space admit a pair of SU(3) structures, which may coincide or not. 

A crucial point is that these two SU(3) structures can be conveniently described in the unifying picture of Hitchin's generalized geometry \cite{HitchinGenCY, GualtieriThesis}, which studies mathematical structures defined on the sum $T\oplus T^*$ of the tangent and cotangent bundle of a manifold. More specifically, the existence of the two SU(3) structures is equivalent to a reduction of the structure group of $TM_6\oplus T^*M_6$ to SU(3)$\times$SU(3). Motivated by the above considerations, we are then led to take this topological fact as a necessary condition for compactifications of type II supergravity to yield an $N=2$ effective action in 4d \cite{GLW1, GLW2}. An appealing approach to the study of general $N=2$ compactifications seems then to assume the existence of an \sst structure as a starting point and then to apply the tools of generalized geometry to study the dimensional reduction.\footnote{A closely related problem to which generalized geometry has been fruitfully applied is the study of supersymmetric flux vacua of type II strings, see e.g. \cite{GMPT1, GMPT2, JeschekWittboth, GeneralizedCalibrations, GMPT3, MicuPaltiTasinato, TomasielloDolbeault, KoerberTsimpis}.}

The study of SU(3)$\times$SU(3) structure compactifications preserving eight supercharges was started in \cite{GLW1} and pursued in \cite{GLW2}. In these papers some relevant terms of the $N=2$ action were obtained. In particular, using Hitchin's results \cite{HitchinGenCY} about the special K\"ahler geometry on the deformation space of generalized structures, \cite{GLW1} studied the SU(3) structure deformations, matching them with the internal metric and $b$--field deformations defining $N=2$ scalar kinetic terms. In \cite{CassaniBilal} we generalized this correspondence to the \sst structure case, also discussing the geometric origin of the period matrices for the $N=2$ special K\"ahler geometry. Furthermore, via a reduction of the gravitino susy transformations, \cite{GLW1, GLW2} derived the $N=2$ Killing prepotentials defining the 4d gaugings. These contain both electric and magnetic charges, originating from the NS, RR, geometric (and possibly non-geometric) background fluxes. The magnetic charges are consistently introduced in a local $N=2$ lagrangian as mass terms for antisymmetric rank--2 tensors \cite{LouisMicu, TheisVandoren, N=2withTensor1, N=2withTensor2, DeWitSamtlebenTrig}. 

A further result in this context is that the $N=1$ supersymmetry conditions obtained from the 10d and 4d approaches to type II vacua admitting \sst structure were shown to be equivalent \cite{KoerberMartucci10to4, CassaniBilal}.

From a purely four dimensional supergravity perspective, \cite{D'AuriaFerrTrigiante} constructed an $N=2$ lagrangian containing the same set of charges appearing in the Killing prepotentials of \cite{GLW2}. In particular, a symplectically invariant and mirror symmetric expression for the $N=2$ scalar potential was obtained.

Despite these results, a complete derivation of the 4d effective action via the dimensional reduction has not appeared in the literature. The purpose of the present paper is to fill this gap for what concerns the bosonic sector.

At this point a very important remark is in order: taken alone, the existence of an \sst structure, though necessary, is far from guaranteeing the 4d theory to exhibit the features of $N=2$ supergravity. Indeed, at a first step most of the results described above were derived working at a point of the internal manifold and preserving all the Kaluza-Klein modes. In order to get a truly four dimensional action one needs to define a mode truncation, and this is done expanding the 10d fields on a finite basis of internal forms. Compatibility with $N=2$ supergravity requires this basis to respect a restrictive set of geometrical constraints, which have been identified in \cite{GLW1, GLW2}, further analysed for SU(3) structure reductions in \cite{MinasianKashani} and revisited in \cite{CassaniBilal}. It is worth saying that in all these studies the dimensional reduction is supposed to proceed similarly to the Calabi-Yau one.

However, already for the strictly SU(3) structure case, it is difficult to exhibit an explicit reduction ansatz. Recently this was achieved in \cite{KashaniPoorNearlyKahler} for the particular SU(3) structure class of nearly K\"ahler manifolds (previous developements can be found in \cite{TomasMirrorSymFl, MinasianKashani}). Another point is that, once a reduction ansatz is identified, it is not guaranteed that the 4d fields defined by the truncation do correspond to (all the) light degrees of freedom. In other words, one should check whether the obtained 4d $N=2$ theory also corresponds to a \emph{low energy} effective theory, and if the truncation captures all the light degrees of freedom associated with the compactification under study.

In this paper we will not address these last issues, also due to their background dependence: indeed the standard Kaluza-Klein procedure identifying the masses of the 4d degrees of freedom passes through the linearization of the equations of motion for fluctuations of the fields around a chosen vacuum. For what concerns the basis forms defining the truncation, we will assume they satisfy the needed constraints, and study the 4d $N=2$ theory as obtained from the dimensional reduction. Furthermore, our analysis is entirely classical and based on the supergravity approximation.\footnote{For the relevance of quantum corrections in this generalized geometry context, see \cite{TorsionSusyBreak}.} 

Here is a summary of the paper and of its results. Our starting point is the `democratic' version of type II supergravities formulated in \cite{Democratics}, which we shortly review in section \ref{DemocraticFormulation}. The RR sector is described by a field strength consisting of a sum of forms of all possible even or odd degrees, and submitted to a self-duality constraint. Because of this homogeneous treatement of the different form degrees, the democratic formulation is particularly suitable for generalized geometry applications (in which context it was first adopted in \cite{GMPT1}).

Section \ref{su3su3summary} recalls the needed notions concerning \sst structures and their deformations, and discusses the basis of expansion forms defining the mode truncation.

Next we approach the type II dimensional reduction, studying the NSNS and RR sectors separately. While the results for the NSNS sector are valid indifferently for IIA and IIB, for what concerns the RR sector we will concentrate on type IIA.

In section \ref{ReductionNSNS} we deal with the reduction of the NSNS sector. We reformulate the different terms in the generalized geometry language, then we implement the truncation ansatz. In particular we focus on the 4d scalar potential: we find and prove a formula expressing the internal NSNS sector in terms of the \sst structure data, and we apply it to derive the scalar potential.

Then in section \ref{ReductionRRsector} we turn to the RR sector. Instead of directly reducing the action, we choose to reduce the equations of motion. Due to the RR self-duality constraint, these can also be read as Bianchi identities. The expansion of the democratic RR field on the internal basis automatically introduces forms of all possible degrees in the 4d spacetime. A subset of the reduced RR equations is interpreted as 4d Bianchi identities, which are solved defining in this way the 4d fundamental fields. The remaining equations are seen as 4d equations of motion, from which we reconstruct the reduced action. The theory we obtain contains massive 2--forms, and is in agreement with the one derived in \cite{D'AuriaFerrTrigiante}. Known results for SU(3) structure compactifications are also recovered.

In section$\:$\ref{discussion} we make some final considerations. We conclude with two appendices: Appendix \ref{conventions} summarizes our conventions, while Appendix \ref{StandardIIAwithFluxes} illustrates the compatibility of the democratic RR equations of motion with the standard type IIA action, including some subtleties related to the presence of background fluxes.

\section{Democratic formulation of type II supergravity}\label{DemocraticFormulation}

\setcounter{equation}{0}

We start with a brief account of some relevant facts concerning the `democratic' formulation of type II supergravities given in \cite{Democratics}. We also took a few notions from \cite{BelovMoore}.

We will just consider the bosonic (NSNS + RR) sector of the theory. The NSNS spectrum consists of the 10d spacetime metric, the 2--form $\hat B$ and the dilaton $\phi$. The corresponding action has the standard (string frame) form\footnote{Here and in the following, the hat symbol denotes ten-dimensional fields (no hat is needed for the dilaton). See Appendix$\:$\ref{conventions} for our other conventions.}
\be\label{eq:NSsector}
S_{\mathrm{NS}} = \frac{1}{2}\int_{M_{10}} e^{-2\phi}\Big( \hat R*1 + 4 d\phi\wedge *d\phi - \frac{1}{2}\hat H\wedge *\hat H\Big)\;.
\ee
The 3--form $\hat H$ is subject to the Bianchi identity 
\be
\label{eq:BianchiForH} d \hat H=0\;,
\ee
which for topologically trivial configurations is globally solved by $\hat H=d \hat B$, while for more general topologies the global solution is
\be\label{eq:SplittingH}
\hat H= \hat H^{\mathrm{fl}} + d\hat B\;,
\ee
where $\hat H^{\mathrm{fl}}$ is a cohomologically non-trivial representative (`fl' stands for `flux'). Notice that this splitting of $\hat H$ allows us to work with globally defined quantities: we could have insisted in writing $\hat H=d\hat B$, but in this case generically the form $\hat B$ wouldn't be globally defined.

We now pass to the RR sector. In the democratic approach to type IIA (IIB), it describes the dynamics of a field ${\bf \hat F}$ consisting of a formal sum of forms of all possible even (odd) degrees:
\be\label{eq:DemocraticRRfield}
{\bf \hat F}= \hat F_0 + \hat F_2 + \ldots + \hat F_{10}\quad\textrm{in IIA}\;,\quad \textrm{while}\qquad {\bf \hat F}= \hat F_1 + \hat F_3 + \ldots + \hat F_{9}   \quad \textrm{in IIB.}
\ee
In order to avoid a doubling of the degrees of freedom with respect to the usual formulation in which only the forms of lower degree appear, a self-duality constraint is imposed on the RR field. In the Hodge-$*$ conventions fixed in appendix$\:$\ref{conventions}, this constraint reads
\be\label{eq:10dSelfDualityF}
\hat{\bf F}= \lambda (* \hat{\bf F})\;,\qquad \textrm{with}\qquad \lambda(\hat F_k) = (-)^{[\frac{k+1}{2}]}\hat F_k\;.
\ee
In the absence of localized sources, the dynamics of the field ${\bf \hat F}$ is described by the following equation of motion (EoM from now on):
\be\label{eq:10dDemocraticRReom/Bianchi}
(d+\hat H\wedge) *\, {\bf \hat F} =0 \qquad \Leftrightarrow\qquad
(d-\hat H\wedge) {\bf \hat F} =0\;,
\ee
where the two expressions are equivalent due to (\ref{eq:10dSelfDualityF}). The second one has the form of a Bianchi identity, and for topologically trivial configurations is globally solved by
\be\label{eq:FintermsOfC}
{\bf \hat F }= (d-\hat H\wedge) {\bf \hat C } + e^{\hat B}\hat F_0\;,
\ee
where ${\bf \hat C}$ is a sum of RR potentials of odd (even) degree for type IIA (IIB), $\hat F_0$ is a constant (present only in type IIA), and $e^{\hat B}\equiv 1+ \hat B\wedge + \textstyle{\frac{1}{2}}\hat B\wedge \hat B\wedge +\ldots\;$. 

Once (\ref{eq:FintermsOfC}) is established, the first expression in (\ref{eq:10dDemocraticRReom/Bianchi}) can be derived by varying the potentials ${\bf \hat C}$ in the following \emph{pseudo}-action \cite{Democratics}:
\be\label{eq:RRpseudoAction}
S_{\mathrm{RR}} \; = \; - \frac{1}{8}\int_{M_{10}} \!\!\big[\,{\bf \hat{F}} \wedge  *{\bf \hat{F}}\,\big]_{10}\;,
\ee
where the notation $[\;]_{10}$ means that we pick the form of maximal degree 10. The prefix `pseudo-' means that (\ref{eq:RRpseudoAction}) contains redundant RR degrees of freedom, and should be considered just as a device to obtain their EoM. The redundancy is then removed at the level of the EoM by the self-duality constraint (\ref{eq:10dSelfDualityF}), which does not descend from (\ref{eq:RRpseudoAction}) and has be imposed by hand. A further peculiarity of this pseudo-action is that it does not contain any Chern-Simons term, which is instead present in the usual formulations of type II supergravities.

A {\it bona fide} action, containing just the independent degrees of freedom, can be recovered by breaking the democracy among the RR differential forms: a half of the $\hat F_k$ has to be eliminated exploiting the self-duality relation. The choice of the forms to keep is not unique, and in some cases the presence of localized sources can suggest the most convenient option \cite{Democratics, VilladZwirner}. In appendix$\;$\ref{StandardIIAwithFluxes} we discuss how the action of standard type IIA supergravity without localized sources can be recovered, also taking into account a deformation of the Chern-Simons term due to background fluxes.

In the following we will also need the EoM for the $\hat B$--field, which is obtained by varying the complete democratic pseudo-action $S_{\mathrm{NS}} + S_{\mathrm{RR}}$. After using the first of (\ref{eq:10dDemocraticRReom/Bianchi}), this reads:
\be\label{eq:EoMfor10dimB}
d( e^{-2 \phi}  *\hat H) -\frac{1}{2} [{\bf \hat F}\wedge * {\bf \hat F}]_8 = 0\;.
\ee

\section{SU(3)$\times$SU(3) structures}\label{su3su3summary}

\setcounter{equation}{0}

\subsection{Supergravity fields from SU(3)$\times$SU(3) structures}\label{SugraFieldsAndsst}

In this section we introduce \sst structures on $T M_6\oplus T^*M_6$, specifying in this way the class of 6d manifolds on which we wish to study general dimensional reductions of type II supergravity. Most of the needed generalized geometry notions have been discussed in our previous work \cite{CassaniBilal}, therefore here we just summarize some fundamentals, together with the necessary formulas. A more extensive review of generalized geometry can be found in \cite{GMPT3}, while for the mathematical details we refer to the original works \cite{HitchinGenCY, GualtieriThesis}.

The bundle $T M_6\oplus T^*M_6$ is naturally endowed with an O(6,6) structure. Reductions of this structure group can be defined starting from Spin(6,6) spinors, which are isomorphically mapped to sections of $\wedge^\bullet T^*M_6 $, i.e. forms of mixed degree (polyforms). In the polyform picture, the Clifford action $\cdot$ on Spin(6,6) spinors is realized by elements of $T\oplus T^*$ acting on $\wedge^\bullet T^*$ as follows: if $X= v + \zeta \in T\oplus T^*$ and $A\in \wedge^\bullet T^*$, then
\be\label{eq:CliffordAction}
X\cdot A = (\iota_{v} + \zeta\wedge)A\;.
\ee
An antisymmetric product between two polyforms $A,B$ is defined via the Mukai pairing:
\be\label{eq:DefMukai}
\langle A,B\rangle\, =\, [\lambda(A)\wedge B]_{6}\;,
\ee
where, as in section$\:$\ref{DemocraticFormulation}, $\,\lambda(A_k) = (-)^{[\frac{k+1}{2}]}A_k$, while $[\;]_{6}$ picks the form of top degree.

The characterization of an SU(3)$\times$SU(3) structure on $TM_6\oplus T^*M_6$ requires a pair of globally defined complex polyforms $\Phi_+$ and $\Phi_-$, sections of $\wedge^{\mathrm{even}}T^*\otimes \mathbb C$ and $\wedge^{\mathrm{odd}}T^*\otimes \mathbb C$ respectively. Both $\Phi_\pm$ have to admit a six-dimensional space of annihilators, i.e. they should be \emph{pure} spinors. Furthermore, they need to satisfy the condition
\be
\label{eq:compatibility}\langle \Phi_+,  X\cdot \Phi_-\rangle \,=\, 0\, =\, \langle \bar \Phi_+, X\cdot \Phi_-\rangle\qquad\quad \forall X\in T\oplus T^*\;.
\ee
Such a pure spinors pair defines a metric $\cl G$ on $T\oplus T^*$. We demand $\cl G$ be positive definite. Then $\Phi_\pm$ are called \emph{compatible}.
Lastly, we require they have nowhere vanishing, equal pairings:
\be
\label{eq:equalNorm}\langle \Phi_+,  \bar\Phi_+\rangle = \langle \Phi_- ,  \bar\Phi_-\rangle \neq 0\;.
\ee

Now, the crucial point for supergravity applications is that the specification of an \sst structure automatically fixes all the NSNS data of the compact space, i.e. it provides a metric $g$, a 2--form $b$ and a dilaton $\phi$ on $M_6$. Moreover, it yields a pair of SU(3) structures for $M_6$, and therefore a pair of globally defined Spin(6) spinors (with positive chirality) $\eta^1_+$ and $\eta^2_+$. Let's see how these data are encoded in the generalized geometry objects.

From $\Phi_\pm$ one can build a pair of commuting \emph{generalized almost complex structures} $\cl J_\pm$, i.e. maps $T\oplus T^*\to T\oplus T^*$ squaring to $-id_{T\oplus T^*}$, via 
\be\label{eq:RelJPhi}
\mathcal J^{\;\Lambda}_{\pm\;\Sigma} = 4i\frac{\langle \re\Phi_\pm , \Gamma^{\Lambda}_{\;\;\Sigma}\re \Phi_\pm \rangle}{\langle \Phi_\pm , \bar\Phi_\pm \rangle}\;,
\ee
where the indices $\Lambda ,\Sigma = 1, \ldots ,12$ run over $T\oplus T^*$, and $\Gamma^{\Lambda\Sigma}$ denotes the antisymmetric product of two Cliff(6,6) gamma matrices. Recalling (\ref{eq:CliffordAction}), at each point of $M_6$ we identify these gamma matrices with the basis elements of $T\oplus T^*$: $\Gamma^\Lambda\;=\; (d y^m\!\wedge \,,\, \iota_{\partial_m})$. The $T\oplus T^*$ indices are lowered with the natural $(6,6)$-signature metric $\mathcal I_{\Lambda\Sigma} = {{0\; 1}\choose {1 \;0}}$ on $T\oplus T^*$, which also enters in $\{\Gamma^\Lambda,\Gamma^\Sigma\}=\mathcal I^{\Lambda\Sigma}$. We remark that $\cl J_\pm$ in (\ref{eq:RelJPhi}) do not depend on the overall phase of $\Phi_\pm$: indeed, since $\langle \Phi , \Gamma^{\Lambda}_{\;\;\Sigma}\Phi \rangle=0$ \cite{HitchinGenCY}, one has $2\langle \re\Phi , \Gamma^{\Lambda}_{\;\;\Sigma}\re \Phi \rangle = \langle \Phi , \Gamma^{\Lambda}_{\;\;\Sigma}\bar \Phi \rangle$.

A metric $\cl G$ on $T\oplus T^*$ is then obtained via $\mathcal G:= -\mathcal J_+ \mathcal J_- = -\mathcal J_- \mathcal J_+$, and it can be shown \cite{GualtieriThesis} that its general form is:
\be\label{eq:TT*metric}
\mathcal G^{\Lambda}_{\;\;\Sigma}=  \cl B \left(\begin{array}{cc} 0 &  g^{-1} \\ [1mm] g & 0   \end{array}\right)\cl B^{-1}\;, \quad\textrm{with} \quad
\cl B= \left(\begin{array}{cc}1 & 0 \\ [1mm] -b & 1 \end{array}\right) \;,
\ee
where $b_{mn}$ is an antisymmetric 2--tensor (to be identified with the NS 2--form), while $g_{mn}$ is a metric for $M_6$, positive definite thanks to the assumed positive-definiteness of $\cl G_{\Lambda\Sigma}$. Taken alone, $\cl G$ defines a reduction of the $T\oplus T^*$ structure group to O(6)$\times$O(6)$\,\subset\,$O(6,6), providing a metric $g$ and a 2--form $b$ on $M_6$. The specification of the commuting pair\footnote{The commuting $\cl J_+,\cl J_-$ defining a positive definite $\cl G$ are called compatible. It can be shown \cite{TomasielloDolbeault} that $[\cl J_+,\cl J_-]= 0$ is equivalent to eq.$\:$(\ref{eq:compatibility}).} $\cl J_+, \cl J_-$ determines a further reduction to U(3)$\times$U(3), and this implies the existence of a pair of U(3) structures for $TM_6$. Indeed, it was shown in \cite{GualtieriThesis} that $\cl J_\pm$ take the form
\be\label{eq:CurlyJ+-}
\mathcal J^{\;\Lambda}_{\pm\;\Sigma} = \frac{1}{2} \cl B         
\left(\begin{array}{cc} -(I_1 \mp I_2) & -(J_1^{-1} \pm J_2^{-1})  \\ [1mm] J_1 \pm J_2 & I^T_1 \mp I^T_2   \end{array}\right)
\cl B^{-1}\;,
\ee
where $(I_k)^{\!m}_{\;\,n}$ and $(J_k)_{mn}$ ($k=1,2$) are respectively an almost complex structure ($I_k:T\to T$ such that $I_k^2=-id$) and an antisymmetric 2--tensor. Each pair $(I_k,J_k)$ identifies an U(3) structure for $TM_6$, and is related to the same metric on $M_6$ via $g_{mn}= J_{mp}I^p_{\;\;n}$.

The further reduction of the $T\oplus T^*$ structure group to \sst amounts to fix the overall phases of $\Phi_+$ and $\Phi_-$ (this also specifies a pair of SU(3) structures inside the U(3) ones), as well as to choose the pure spinor normalization, on which the previous definitions of $\cl J_\pm$ and $\cl G$ do not depend. Recalling (\ref{eq:equalNorm}), the norm of $\Phi_\pm$  corresponds to a single positive function over $M_6$, which we relate to the dilaton. More precisely, denoting as $vol_6$ the volume form on $M_6$, we take:
\be\label{eq:PureSpNorm}
||\Phi_\pm||^2vol_6 := i\langle \Phi_\pm , \bar \Phi_\pm  \rangle = 8e^{-2\phi}vol_6\;.
\ee

To each pair $(I_k,J_k)$, $k=1,2$, is associated an SU(3)--invariant globally defined Spin(6) spinor with positive chirality $\eta^k_+$ (see subsect.$\:$\ref{SU3strConv} of the appendix for further details on the relation between SU(3)--invariant spinors and tensors). An explicit relation between the Spin(6) spinors $\eta^1_+$ and $\eta^2_+$ and the Spin(6,6) pure spinors defining \sst structures with vanishing $b$--field (call them $\Phi^0_\pm$) is established by\footnote{Further developements on explicit constructions of compatible pure spinor pairs can be found in \cite{HalmagyiTomasiello}.} \cite{GMPT2}:
\be\label{eq:defPhi0}
\slashPhipm = 8\eta^1_+\otimes \eta^{2\dag}_\pm\;,
\ee
where the action of the Clifford map ``$\,/\,$'' is: 
\be\label{eq:DefCliffordMap}
/\,:\; dy^{m_1}\wedge\ldots \wedge dy^{m_k}\;\mapsto\; \gamma^{m_1\ldots m_k}\;,
\ee
while to evaluate the bispinor in the rhs of (\ref{eq:defPhi0}) the Fierz identity (\ref{eq:fierz}) is used. We identify the product of the two nonvanishing Spin(6) spinor norms with the dilaton:
\be\label{eq:ProductNormsIsDilaton}
||\eta^1_\pm|| \,||\eta^2_\pm|| = e^{-\phi}\;,
\ee
so that (\ref{eq:PureSpNorm}) is ensured by (\ref{eq:PSnorm}). \sst structures with nonvanishing $b$ can then be recovered via the following $b$--transform on $\Phi^0_\pm$:
\be\label{eq:bTransform}
\Phi_{\pm} = e^{-b}\Phi^0_\pm\;.
\ee

This `bispinor picture', in which $\Phi^0_\pm$ are treated as in (\ref{eq:defPhi0}), is often advantageous in concrete computations. Some more technical details are reported in subsection$\:$\ref{sstAndSpin6} of the appendix. In particular, (\ref{eq:BasisDiamond}) provides an explicit basis for the decomposition of the elements of $\wedge^\bullet T^*$ in representations of \sst$\!\!$, while eq.$\:$(\ref{eq:MukaiUnderClifford}) illustrates how to evaluate the Mukai pairing in this picture.

The two Spin(6) spinors $\eta^1_+$ and $\eta^2_+$ provided by the \sst structure are precisely the internal spinors to be used in the Spin(9,1)$\,\to\,$Spin(3,1)$\times$ Spin(6) decomposition of the two type II supersymmetry parameters we mentioned in the introduction. Choosing to 
reduce the first 10d susy parameter employing just $\eta^1_\pm$, and the second using just $\eta^2_\pm$, yields a decomposition ansatz preserving eight supercharges, and therefore $N=2$ in 4d.

Finally, we remark that the two SU(3) structures defined by the \sst structure on $T\oplus T^*$ may also be the same. In this case the internal manifold $M_6$ has a strictly SU(3) structure, and the spinors $\eta^1_+$ and $\eta^2_+$ are everywhere parallel. However, generically we will consider the two spinors being independent almost everywhere, and becoming parallel at some points: in this situation a local SU(2) structure for $TM_6$ is defined, but not a global one. Nowhere parallel $\eta_+^1$ and $\eta_+^2$ identify a global SU(2) structure; this last case is rather associated with $N=4$ compactifications since each of the 10d susy parameters can be decomposed on either $\eta^1_+$ and $\eta^2_+$ \cite{TorsionSusyBreak}.

\subsection{Deformations of \sst structures}\label{sstDeformations}

Compactifying on a given class of manifolds requires knowledge of the corresponding moduli space. Indeed, the moduli associated with the internal metric deformations constitute scalar fields of the compactified theory, and their kinetic terms are specified by the metric on the space of deformations. In the following we resume the description of \sst structure deformations given in \cite{CassaniBilal}, adding some further specifications. Other physical applications concerning the deformation theory of generalized structures developed in \cite{HitchinGenCY, GualtieriThesis, LiDeformGenCY} can be found in \cite{TomasielloDolbeault, KoerberMartucci10to4, HalmagyiTomasiello}.

In the notation of \cite{CassaniBilal}, we write small deformations of the pure spinors $\Phi_+$ and $\Phi_-$ as:
\be\label{eq:variationPhi}
\delta \Phi_\pm = \delta \kappa_\pm \Phi_\pm + \delta_{\mathrm{tr}}\Phi_\pm +\delta \chi_\pm\;. 
\ee
Because of condition (\ref{eq:equalNorm}), the real parts of the scalars $\delta \kappa_\pm$ need to be equal (the imaginary parts are instead independent).
The independent complex deformations $\delta\chi_-$ and $\delta\chi_+$, being sections respectively of the $U_{{\bf \bar 3},{\bf \bar 3}}$ and $ U_{{\bf \bar 3},{\bf 3}}\,$ bundles defined in subsection$\:$\ref{sstAndSpin6} of the appendix, at each point of $M_6$ can be parameterized using the basis (\ref{eq:BasisDiamond}) as 
\be
\delta\chi_\pm = e^{-b}\delta\chi^0_\pm\;\;,\qquad \textrm{with}\qquad \delta\chi^0_+ = (\delta \chi_+)_{\bar\imath_1 j_2} \gamma^{\bar\imath_1} \Phi^0_+ \gamma ^{j_2} \quad,\quad \delta\chi^0_- = (\delta \chi_-)_{\bar\imath_1 \bar\jmath_2} \gamma^{\bar\imath_1} \Phi^0_- \gamma ^{\bar\jmath_2}\;.
\ee
Here and in the following the indices $\bar\imath_1, i_1$ are (anti)holomorphic with respect to the almost complex structure $I_1$, and analogously for $\bar\jmath_2, j_2$ with respect to $I_2$. The complex tensors $(\delta\chi_+)_{mn}$ and $(\delta \chi_-)_{mn}$ satisfy
\be\label{eq:projectorsDeltaChi+-}
\bar P_{\!1m}^{\:\;\;p} P_{\!2n}^{\:\;\;q} (\delta \chi_+)_{pq} = (\delta \chi_+)_{mn}\qquad,\qquad \bar P_{\!1m}^{\:\;\;p} \bar P_{\!2n}^{\:\;\;q} (\delta \chi_-)_{pq} = (\delta \chi_-)_{mn}\;,
\ee
where $(P_k)_{m}^{\;\;\,n} = \frac{1}{2}(\delta - iI_k)_m^{\;\;n}$ is the holomorphic projector associated with $I_k\,, \;k=1,2$.

With respect to \cite{CassaniBilal}, in (\ref{eq:variationPhi}) we have also considered possible additional deformations $\delta_{\mathrm{tr}} \Phi_+$ and $\delta_{\mathrm{tr}} \Phi_-$ living in the \sst `triplets' $({\bf 3},{\bf 1}) \oplus ({\bf {\bar 3}},{\bf 1})\oplus ({\bf 1},{\bf 3}) \oplus ({\bf 1},{\bf{\bar 3}})$. These are precisely the pure spinor deformations constrained by the compatibility condition (\ref{eq:compatibility}), which requires them to be performed simultaneously. More specifically, using the basis (\ref{eq:BasisDiamond}), at a point of $M_6$ a parameterization of these simultaneous variations is
\be
\label{eq:DeformationsInTriplets} \delta_{\mathrm{tr}} \Phi_+ = e^{-b}\big(\delta u_{i_1}\gamma^{i_1}\bar\Phi^0_- + \delta \bar v_{\bar\imath_2} \Phi^0_- \gamma^{\bar \imath_2} \big)\quad,\quad
\delta_{\mathrm{tr}} \Phi_- = - e^{-b}\big(\delta u_{i_1}\gamma^{i_1}\bar\Phi^0_+ +  \delta  v_{i_2} \Phi^0_+ \gamma^{i_2}\big)\;,
\ee
where
\be
\delta u_{i_1} = \frac{1}{2}(\delta - i I_1)_{i_1}^{\;\;m} \delta u_m\qquad,\qquad \delta v_{i_2} = \frac{1}{2}(\delta - i I_2)_{i_2}^{\;\;m} \delta v_m\;,
\ee
$\delta u_m$ and $\delta v_m$ being real and independent small parameters. Via the Clifford map, expression (\ref{eq:DeformationsInTriplets}) can be read either in the bispinor picture, or in the polyform picture (in this case $\buildrel\to\over\gamma\!{}^{i_1}$ and $\buildrel\leftarrow\over\gamma\!{}^{i_2}$ are mapped to elements of $(T\oplus T^*)\otimes \mathbb C$, see subsection$\:$\ref{sstAndSpin6} of the appendix). 

The pure spinor deformations induce deformations of the associated generalized almost complex structure. Recalling (\ref{eq:RelJPhi}), for both even/odd parities the relation is given by:
\be\label{eq:VarJ}
\delta\mathcal J_{\Lambda\Sigma} = 8i\frac{\langle \re( \delta_{\mathrm{tr}}\Phi + \delta\chi) , \Gamma_{\Lambda\Sigma}\re\Phi \rangle}{\langle \Phi , \bar \Phi \rangle}\;.
\ee
Notice that the rescaling piece $\delta\kappa$ drops.

In \cite{HitchinGenCY, HitchinGeom3forms} Hitchin shows that the space of even (odd) pure spinor deformations {\it at a point of $M_6$} admits a rigid special K\"ahler metric, and that a local special K\"ahler geometry can be obtained by taking the quotient with the $\mathbb C^*$ action generated by rescalings of $\Phi_+$ ($\Phi_-$) (see \cite{GLW1} for a detailed review). This quotient coincides with the deformation space of the associated generalized almost complex structure $\cl J_+$ ($\cl J_-$). The corresponding K\"ahler potentials $K_\pm$ are the Hitchin functions 
\be\label{eq:KpmNotIntegr}
e^{-K_{\pm}}= i\langle \Phi_\pm,\bar\Phi_\pm \rangle\;.
\ee
Varying $K_\pm$ with respect to the holomorphic/antiholomorphic pure spinor deformations as done in \cite{CassaniBilal}, but this time taking also $\delta_{\mathrm{tr}} \Phi_\pm$ in (\ref{eq:DeformationsInTriplets}) into account, yields the metric on the space of compatible $\cl J_+, \cl J_-\,$, i.e. on the space of U(3)$\times$U(3) structures (at a point of $M_6$):
\be\label{eq:MetricSpaceU3U3}
ds^2 = \delta^{\mathrm{holo}}\delta^{\mathrm{anti}}(K_+ + K_-) =  - \frac{\langle \delta\chi_-, \delta \bar\chi_-\rangle}{\langle \Phi_-, \bar\Phi_-\rangle} - \frac{\langle \delta\chi_+, \delta \bar\chi_+\rangle}{\langle \Phi_+, \bar\Phi_+\rangle} + 2 g^{mn}(\delta u_m \delta u_n + \delta v_m\delta v_n)\;,
\ee
where the last term arises from the equal contributions (the computation uses (\ref{eq:MukaiUnderClifford})):
\be
- \frac{\langle \delta_{\mathrm{tr}} \Phi_\pm, \delta_{\mathrm{tr}} \bar \Phi_\pm\rangle}{\langle \Phi_\pm, \bar\Phi_\pm\rangle} = g^{mn}(\delta u_m \delta u_n + \delta v_m\delta v_n)\;.
\ee
Since $\delta_{\mathrm{tr}}{\Phi_-}$ and $\delta_{\mathrm{tr}}{\Phi_+}$ are not independent, the space of U(3)$\times$U(3) structures with the metric (\ref{eq:MetricSpaceU3U3}) is not a direct product of ${\cal J_+}$ and ${\cal J_-}$ deformation spaces.

\subsection{Truncating to a finite set of modes}\label{DefiningTruncation}

In order to dimensionally reduce the higher dimensional supergravity, one has to truncate the modes of the 10d fields along $M_6$ to a finite set. Such a truncation ansatz can be specified providing a basis of internal differential forms on which to expand the 10d fields. In this paper we are interested in general \sst structure reductions leading to $N=2$ supergravities in four dimensions: the requirements needed for this to be achieved were given in \cite{GLW1, GLW2} and, for SU(3) structure reductions, carefully scrutinized in \cite{MinasianKashani}. In \cite{CassaniBilal} we partially extended this last analysis to the \sst structure context. In this subsection we collect the relations we will need in order to derive the general form of the four dimensional action. We emphasize that the list we provide is not fully exhaustive.

A first condition for a standard 4d, $N=2$ action requires to truncate all the 10d field components transforming in the $({\bf 3},{\bf 1}) \oplus ({\bf {\bar 3}},{\bf 1})\oplus ({\bf 1},{\bf 3}) \oplus ({\bf 1},{\bf{\bar 3}})$ representation of \hbox{\sst\!\!:} indeed these would assemble in non-standard 4d spin-3/2 multiplets \cite{GLW2, GLW1}. This requirement concerns in particular the pure spinor deformations $\delta_{\mathrm{tr}} \Phi_\pm$ introduced in subsect.$\:$\ref{sstDeformations}. After the truncation of $\delta_{\mathrm{tr}}{\Phi_\pm}$, the space of U(3)$\times$U(3) structures splits in a direct product describing the independent deformations of ${\cal J_+}$ and ${\cal J_-}$. As we will see in subsection$\:$\ref{ScalarKinTerms}, this space coincides with the deformation space of the generalized metric $\cl G$, i.e. of the internal metric and $b$--field. Notice the similarity with the Calabi--Yau case, where the (finite-dimensional) moduli space splits in the product of two local special K\"ahler manifolds describing the independent complex-- and K\"ahler--structure deformations \cite{CandelasOssa}. From the point of view of 4d $N=2$ supergravity, these two sets of Calabi--Yau moduli define the scalar components of the vector multiplets and a subset of the scalar components of the hypermultiplets. The special K\"ahler structure of the vector multiplet scalar manifold is necessary in order to have consistency with the $N=2$ supergravity formalism, while the special K\"ahler manifold associated with the hypersector constitutes the basis of a {\it special} quaternionic manifold \cite{FerraraSabharwal}. 

In the general \sst structure dimensional reduction, several requirements on the expansion forms are needed in order to ensure that the local special K\"ahler structure on the (now independent) spaces of $\cl J_-$ and of $\cl J_+$ deformations at a point of $M_6$ be inherited by the finite-dimensional spaces of 4d fields identified by the truncation. We call these spaces $\mathscr M_-$ and $\mathscr M_+$ respectively, with $\mathrm{dim}\mathscr M_\pm = b^{\pm}$. 

In order to preserve the symplectic structure defined by the Mukai pairing, these real basis forms should arrange in symplectic vectors $\Sigma_\pm$:
\be\label{eq:Sigma+-AsVectors}
\Sigma_+^{\mathbb A} = \left(\begin{array}{c} \tilde\om^A \\ \om_A \end{array}\right)\qquad , \qquad \Sigma_-^{\mathbb I} = \left(\begin{array}{c} \beta^I \\ \alpha_I \end{array}\right)\;,
\ee
where $\Sigma_+$ contains even forms, while $\Sigma_-$ is made of odd forms. A main point is that these forms need not be of pure degree, i.e. are in general polyforms. The range of the indices is: $A,B=0,1,\ldots,b^+$ and $I,J\,=\,0,1,\ldots, b^-$. We also introduce the symplectic indices $\mathbb A, \mathbb B= 1,2,\ldots, 2(b^+ + 1)$ and $\mathbb I, \mathbb J= 1,2,\ldots, 2(b^- + 1)$. The pairings of the basis forms are then required to satisfy:
\be
\int_{M_6}\langle \Sigma_+^{\mathbb A} , \Sigma_+^{\mathbb B} \rangle = (\mathbb{S_+})^{-1\,\mathbb{AB}}\qquad,\qquad \int_{M_6}\langle \Sigma_-^{\mathbb I} , \Sigma_-^{\mathbb J} \rangle = (\mathbb{S_-})^{-1\,\mathbb{IJ}}\;,
\ee
where $\mathbb S_\pm = {\,0\;\,\,1 \choose {-1\;0} }$ are the symplectic metrics of Sp$(2b^\pm +2, \mathbb R)$.

The finite set of modes of the NSNS supergravity fields is specified by defining the expansion of the pure spinors $\Phi_\pm$ determining the \sst structure:
\be\label{eq:expansPhi+-}
\Phi_+ = X^A\om_A - {\cal F}_A \tilde \om^A \qquad, \qquad \Phi_- = Z^I \alpha_I - {\cal G}_I \beta^I\;.
\ee
The complex variables $X^A$ and $Z^I$ represent projective coordinates for the local special K\"ahler manifolds $\mathscr M_+$ and $\mathscr M_-$ respectively, and depend on the 4d spacetime coordinates only. Furthermore $\cl F_A = \cl F_A(X)$, while $\cl G_I= \cl G_I(Z)$. Together these arrange in the symplectic vectors
\be
X^{\mathbb A} = \left(\begin{array}{c} X^A \\ \cl F_A \end{array}\right)\qquad , \qquad Z^{\mathbb I} = \left(\begin{array}{c} Z^I \\ \cl G_I \end{array}\right)\;.
\ee
Then the K\"ahler potentials (\ref{eq:KpmNotIntegr}), now integrated,
\be\label{eq:Kpm}
K_\pm = -\log i\int\langle \Phi_\pm, \bar\Phi_\pm \rangle
\ee
take the standard form of special geometry: for instance $K_+ = -\log i(\bar X^A\cl{F_A}- X^A \bar{\cl F}_A)$.

We remark that in general the expansion forms are moduli-dependent (see \cite{MinasianKashani} for a discussion on this point). However, we assume their derivative with respect to the geometric moduli vanishes in the integrated symplectic pairing \cite{GLW2, CassaniBilal}. 

A further condition that seems necessary for the dimensional reduction to proceed analogously to the Calabi-Yau case is that the ratios
\be\label{eq:RatiosConstant}
\frac{\langle \Sigma_+^{\mathbb A} , \Phi_+ \rangle}{\langle \Phi_+, \bar \Phi_+ \rangle} \quad\textrm{and}\quad \frac{\langle \Sigma_-^{\mathbb I} , \Phi_- \rangle}{\langle \Phi_-, \bar \Phi_- \rangle} \quad \textrm{ be constant on $M_6$.}
\ee
Provided that $\langle Z^J\partial_{Z^I}\alpha_J -\cl G_J\partial_{Z^I}\beta^J,\bar\Phi\rangle=0$  (and similarly for the even basis), this is equivalent to demand that \cite{CassaniBilal}
\be\label{eq:KappaConstant}
\kappa^+_A = \frac{\langle\partial_{X^A}\Phi_+ , \bar\Phi_+ \rangle}{\langle\Phi_+, \bar \Phi_+\rangle} \quad\textrm{and}\quad \kappa^-_I = \frac{\langle\partial_{Z^I}\Phi_- , \bar\Phi_- \rangle}{\langle\Phi_-, \bar \Phi_-\rangle} \quad \textrm{ be constant on $M_6$,}
\ee
where, with reference to (\ref{eq:variationPhi}), $\kappa^-_I$ is such that $\delta \kappa_- = \kappa^-_I\delta Z^I$ (and similarly for $\kappa^+_A$). Conditions (\ref{eq:RatiosConstant}) and (\ref{eq:KappaConstant}) are satisfied when $M_6$ is a Calabi-Yau three-fold. To verify (\ref{eq:RatiosConstant}) one should recall that in the Calabi-Yau case the basis forms in $\Sigma_\pm$ are harmonic and the pure spinors take the SU(3) structure form $\Phi_+ = e^{-\phi}e^{-b-iJ}$ and $\Phi_- =-ie^{-\phi}\Om$ (see (\ref{eq:*Jand*1})). Here $J$ is the K\"ahler form of the Calabi-Yau, $\Om$ is the holomorphic $(3,0)$ form, and the dilaton $\phi$ is constant along $M_6$. For instance, for the harmonic (1,1)--forms $\om_a$ the first expression in (\ref{eq:RatiosConstant}) reads:
\be\label{eq:constancyOfOmContrJ}
3 \frac{\om_a \wedge J\wedge J}{J\wedge J\wedge J} = \om_a \lrcorner J \;,
\ee
where eqs.$\:$(\ref{eq:*Jand*1}) and (\ref{eq:*becomesContract}) were used. Now, harmonicity of $\om_a$ implies $\partial_m(\om_a\lrcorner J) = 0\;$ \cite{Strominger-Yukawa}. In the general \sst structure case (\ref{eq:RatiosConstant}) and (\ref{eq:KappaConstant}) are non-trivial assumptions, and we are going to employ them at several points of the dimensional reduction.

In \cite{CassaniBilal} we also discussed the geometric origin of the period matrices $\cl N_{AB}$ and $\cl M_{IJ}$ associated with the special K\"ahler structure of $\mathscr M_+$ and $\mathscr M_-$ respectively. These matrices were related with the action on the basis polyforms of the 6d $b$-twisted Hodge dual:
\be\label{eq:Def*b}
*_b:=e^{-b}*\lambda e^b\;.
\ee
We introduced the matrices:
\be\label{eq:mathbbMandmathbbN}
\mathbb N^{\mathbb A}_{\;\;\mathbb B} := \int \langle\, \Sigma_+^{\mathbb A},*_b \Sigma_{+\mathbb B} \,\rangle
\qquad,\qquad \mathbb M^{\mathbb I}_{\;\;\mathbb J}:=\int \langle \,\Sigma_-^{\mathbb I},*_b \Sigma_{-\mathbb J}\, \rangle\;,
\ee
where $\Sigma_{+\mathbb B}= \mathbb S_{+\mathbb{BC}}\Sigma_+^{\mathbb C}$, and $\Sigma_{-\mathbb J}= \mathbb S_{-\mathbb{JK}}\Sigma_-^{\mathbb K}$. Also using assumption (\ref{eq:KappaConstant}), we arrived at the result:\footnote{There is an irrelevant global minus sign with respect to \cite{CassaniBilal}, due to a change in the definition of the Mukai pairing.}
\be\label{eq:tildeN}
\widetilde{\mathbb N}:=-\mathbb S_+\mathbb{N}\;=\;\left(\begin{array}{cc} 1 & -\re\mathcal N \\ [1mm] 0 & 1 \end{array}\right)
\left(\begin{array}{cc} \im \mathcal N & 0 \\ [1mm] 0 & (\im \mathcal N)^{-1} \end{array}\right)
\left(\begin{array}{cc} 1 & 0 \\ [1mm] -\re\mathcal N & 1 \end{array}\right)\;,
\ee
together with an identical expression for $\widetilde{\mathbb M}:=-\mathbb S_-\mathbb{M}$, having the period matrix $\cl M$ at the place of $\cl N$. It can be deduced from (\ref{eq:mathbbMandmathbbN}) that the matrices $\widetilde{\mathbb{N}}$ and $\widetilde{\mathbb{M}}$ are symmetric and negative definite. To see that $\widetilde{\mathbb{N}}$ is negative definite it is sufficient to notice that
\be
-(\im \cl N)^{-1\, AB}= \int \langle \tilde\om^A, *_b \tilde\om^B \rangle = \int\langle e^b\tilde\om^A, *\lambda (e^b\tilde\om^B) \rangle = \sum_k \int (e^b\tilde\om^A)_k\lrcorner (e^b\tilde\om^B)_k vol_6\,,
\ee
where $k$ denotes the different form degrees of the polyform $e^b\tilde \om^A$. The argument for $\widetilde{\mathbb{M}}$ is completely analogous. This result concerning the action of the $*_b$ operator generalizes the well-known expression for the usual Hodge $*$ acting on the Calabi-Yau harmonic 3--forms \cite{Suzuki, CeresoleEtAl1}.

An important property of the basis polyforms in $\Sigma_\pm$ is that they need not be closed. Introducing an exterior derivative twisted by the harmonic piece $H^{\mathrm{fl}}$ of the internal \hbox{NS 3-form}
\be
d_{H^{\mathrm{fl}}} = d - H^{\mathrm{fl}}\wedge\;,
\ee
we assume $\Sigma_\pm$ satisfy the differential conditions \cite{GLW2}:
\be\label{eq:AnsatzDerForms}
d_{H^{\mathrm{fl}}} \Sigma_- \sim \mathbb Q \Sigma_+\qquad,\qquad d_{H^{\mathrm{fl}}} \Sigma_+ \sim  \widetilde{\mathbb Q} \Sigma_-\;,\ee
where the symbol $\sim$ means equality up to terms vanishing inside the symplectic pairing,
and $\mathbb Q$ is a $(2b^-+2)\times (2b^+ + 2)$ rectangular matrix of constant parameters encoding the NSNS ($H^{\mathrm{fl}}$ and geometric) fluxes:\footnote{We remark that, as illustrated in \cite{GLW2}, the action of the differential operator $d_{H^{\mathrm{fl}}}$ cannot realize all the possible charges in $\mathbb Q$. This can be achieved only on a non-geometric background, performing the extension $d_{H^{\mathrm{fl}}}\to \cl D$, where $\cal D$ is an operator encoding both geometric and non-geometric fluxes, first introduced in \cite{SheltonTaylorWechtVacua}. Even if we are not concerned with non-geometric backgrounds here, we find it advantageous to employ the general symplectically covariant form of $\mathbb Q$.}
\be
\label{eq:ChargeMatrixQ} \mathbb Q^{\mathbb I}_{\;\;\mathbb A} :=\left(\begin{array}{cc} m^I_{\;\;A} &  q^{IA} \\ e_{IA} & p_I^{\;\;A}   \end{array}\right)\;.
\ee
The matrix $\widetilde{\mathbb Q}$ is simply related to $\mathbb Q$: indeed, since $\int \langle d_{H^{\mathrm{fl}}}\Sigma_-,\Sigma_+ \rangle = \int \langle \Sigma_-, d_{H^{\mathrm{fl}}}\Sigma_+ \rangle$, one has
\be\label{eq:ChargeMatrixTildeQ}
\widetilde{\mathbb Q} = (\mathbb S_+)^{-1}\mathbb Q^T \mathbb S_-\;.
\ee
The nilpotency $(d_{H^{\mathrm{fl}}})^2=0$ implies the quadratic constraints:
\be\label{eq:QuadrConstrSympl}
\mathbb Q (\mathbb S_+)^{-1}\mathbb Q^T\, = \,0\, =\, \mathbb Q^T\mathbb S_- \mathbb Q\,.
\ee

\section{Reduction of the NSNS sector}\label{ReductionNSNS}

\setcounter{equation}{0}

We now apply the notions introduced in the previous section to the dimensional reduction of type II supergravity, starting from the NSNS sector. We assume a background topology of the type $M_{10}= M_{4}\times M_{6}$, where $M_4$ is the 4d `external' spacetime and $M_6$ is a 6d `internal' compact manifold admitting SU(3)$\times$SU(3) structure on $T\oplus T^*$. Coordinates along $M_4$ and $M_6$ are denoted by $x^\mu$ and $y^m$ respectively.

Next we introduce a reduction ansatz for the NSNS fields. For the metric we take
\be\label{eq:10dMetricAnsatz}
d s^2 =  g_{\mu\nu}(x)dx^\mu dx^\nu + g_{mn}(x,y)dy^mdy^n\;.
\ee
The NS 3--form $\hat H$ splits as in (\ref{eq:SplittingH}). The cohomologically non-trivial part has just internal indices: $\hat H^{\mathrm{fl}} \equiv H^{\mathrm{fl}}$, while for the potential $\hat B$ we take
\be
\hat B= B + b\;\;,\quad \textrm{with} \;\;B={\textstyle\frac{1}{2}}B_{\mu\nu}(x)dx^\mu\wedge dx^\nu\quad\textrm{and}  \quad b={\textstyle\frac{1}{2}} b_{mn}(x,y)dy^m\wedge dy^n\;.
\ee 
Finally, we allow a possible dependence of the 10d dilaton on both external and internal coordinates: 
\be\label{eq:10dDilaton}
\phi = \phi(x,y)\;.
\ee 

The absence of the terms with mixed indices $g_{\mu n}$ and $B_{\mu n}$ in the reduction ansatz is a well-known feature of Calabi-Yau compactifications: massless 4d fields from these terms would be in correspondence with covariantly constant vectors on the compact Ricci-flat manifold, which are forbidden by SU(3) holonomy. In the general SU(3) and \sst structure context a motivation for not to include $g_{\mu n}$ and $B_{\mu n}$ in the truncation ansatz was given in \cite{GLW1, GLW2} by observing that these fields transform in the `triplets' of \sst (see the discussion in subsection$\:$\ref{DefiningTruncation} above). Therefore, as in the Calabi-Yau case, the NSNS sector will provide no 4d vectors: these will instead descend from the RR sector.

One can now plug ansatz (\ref{eq:10dMetricAnsatz})--(\ref{eq:10dDilaton}) in (\ref{eq:NSsector}) and derive the NSNS sector decomposition. The treatment of the quadratic terms in the dilaton $\phi$ and NS 3--form $\hat H$ appearing in (\ref{eq:NSsector}) being straightforward, we just have to focus on the Einstein-Hilbert term in the action. Under (\ref{eq:10dMetricAnsatz}) the higher dimensional Ricci scalar becomes
\be\label{eq:DecompR}
\hat R_{10} = R_4 + R_6 -\frac{1}{4}g^{mp}g^{nq}\big( \partial_\mu g_{mp}\partial^\mu g_{nq} - 3 \partial_\mu g_{mn}\partial^\mu g_{pq}\big) -g^{mn}\nabla_{\!4}^2 g_{mn}\;,
\ee
where $R_4$ and $R_6$ are the Ricci scalars associated with the metrics on $M_4$ and $M_6$ respectively, while $\nabla_{\!4}^2$ is the laplacian on $M_4$. One now proceeds in two steps. First substitute (\ref{eq:DecompR}) in $\frac{1}{2}\int_{M_{10}}vol_{10}e^{-2\phi} \hat R_{10}$ and perform the integration by parts ($vol_d$ is the volume form on $M_d$):
\be
-\frac{1}{2}\int_{M_4}vol_4\int_{M_6}vol_6 e^{-2\phi}g^{mn}\nabla_{\!4}^2 g_{mn} \;=\;  \frac{1}{2}\int_{M_4}vol_4\int_{M_6}\partial_\mu(vol_6 e^{-2\phi}g^{mn})\partial^\mu g_{mn}\;.
\ee
Secondly, pass to the 4d Einstein frame by introducing the 4d Weyl rescaled metric (no rescaling is instead performed on the 6d metric):
\be\label{eq:rescale4dmetric}
g_{\mu\nu}^{\mathrm{new}} :=  e^{-2\varphi}g_{\mu\nu}^{\mathrm{old}}\;,
\ee
where the 4d dilaton $\varphi$ is defined as
\be\label{eq:4dDilaton}
e^{-2\varphi}:= \int_{M_6} vol_6 e^{-2\phi}\;.
\ee
Under this rescaling, $R_4^{\mathrm{old}} = e^{-2\varphi}(R_4^{\mathrm{new}} - 6\nabla_{\!4}^2 \varphi - 6 \partial_\mu\varphi \partial^\mu\varphi)$, where on the rhs the indices are raised with the new metric. 

Putting everything together, the reduction of (\ref{eq:NSsector}) results then in:
\begin{eqnarray}
\nnb S_{\mathrm{NS}} 
&=&\; \frac{1}{2}\int_{M_4} vol_{4} \big( R_{4} - 2 \partial_\mu\varphi \partial^\mu\varphi -\frac{1}{12}e^{-4\varphi}H_{\mu\nu\rho}H^{\mu\nu\rho}  \big)  \\
\nnb &&\!\!\!\!\! -\;  \frac{1}{8}\int_{M_4} vol_{4} e^{2\varphi} \int_{M_6} vol_6 e^{-2\phi} g^{mp}g^{nq}\big( \partial_\mu g_{mn}\partial^\mu g_{pq} + \partial_\mu b_{mn}\partial^\mu b_{pq} \big) \\
\nnb &&\!\!\!\!\! -\; \frac{1}{2}\int_{M_4} vol_{4} e^{2\varphi} \int_{M_6} vol_6 e^{-2\phi} \nabla_{\!4}^2 \log \big(e^{-2\phi}\sqrt{g_6}\big)\\
\label{eq:decompositionNSsector}&&\!\!\!\!\! - \;  \int_{M_4} vol_{4} \cl V_{\mathrm{NS}}\;,
\end{eqnarray}
where $g_6 \equiv \det (g_{mn})$ and $\cl V_{\mathrm{NS}}$ is identified with the part of the reduced NSNS sector not containing any 4d spacetime derivative:
\be\label{eq:DefV_NS}
\cl V_{\mathrm{NS}} \equiv -\frac{e^{4\varphi} }{2}\int_{M_6} vol_6 e^{-2\phi} \big( R_6 +4 \partial_m\phi \partial^m\phi -\frac{1}{12} H_{mnp}H^{mnp}  \big)\;,
\ee
and therefore represents the contribution of the NSNS sector to the 4d scalar potential.\footnote{A further contribution to the scalar potential is generated from the RR sector and will be derived in the next section. The total potential of the effective theory will be $\cl V= \cl V_{\mathrm{NS}} + \cl V_{\mathrm{RR}}$.}

The first line of (\ref{eq:decompositionNSsector}) already contains 4d fields only, and is compatible with 4d \hbox{$N=2$} supergravity. In standard fluxless Calabi-Yau compactifications the four dimensional \hbox{$B$--field} is usually dualized to an axion which, together with the 4d dilaton $\varphi$ and two further scalars from the RR sector, defines the bosonic part of the so called universal hypermultiplet. However, as first observed in \cite{LouisMicu}, in the presence of RR magnetic fluxes the NS 2--form acquires mass terms and therefore cannot be dualized to a scalar. Anyway, as shown in \cite{TheisVandoren, N=2withTensor1, N=2withTensor2}, (massive) antisymmetric 2--tensors can be included consistently in an $N=2$ supergravity action. We will have more to say about this in section$\:$\ref{ReductionRRsector}.

The subsequent lines in (\ref{eq:decompositionNSsector}) still need to be reformulated in terms of a truncated set of modes of the fields $g_{mn}, b_{mn}$ and $\phi$. For this purpose, in the forthcoming subsections first we translate these expressions in the language of generalized geometry, relating them with the \sst structure data. Then we implement the expansion in terms of the truncated set of modes introduced in subsection \ref{DefiningTruncation}.

Before discussing the relation with \sst structures, let's briefly recall how the dimensional reduction proceeds when performed on Calabi-Yau manifolds in the absence of background fluxes \cite{BodnerCadavidFerrara}. The Calabi-Yau metric and $b$-field deformations are expressed in terms of harmonic forms, and this also corresponds to the Kaluza-Klein prescription for massless 4d scalars. The second line of (\ref{eq:decompositionNSsector}) can be reformulated as a $\sigma$--model whose metric splits in the sum of the special K\"ahler metrics on the spaces of complex-- and K\"ahler--structure deformations \cite{CandelasOssa}. This yields the kinetic terms for the scalars in the vector multiplets as well as the kinetic terms for a subset of the scalars in the hypermultiplets. 

The last two lines of (\ref{eq:decompositionNSsector}) vanish in Calabi-Yau dimensional reductions. The line \hbox{involving} $\nabla_{\!4}^2 \log \big(e^{-2\phi}\sqrt{g_6}\big)$ vanishes thanks to the internal coordinate independence of this last term: passing it out the integral over $M_6$ and recalling (\ref{eq:4dDilaton}), one is left with the integral over $M_4$ of a total derivative. The constancy of $\nabla_{\!4}^2 \log \sqrt{g_6}$ along the Calabi-Yau can be seen as follows. Recall that $\sqrt{g_6}$ depends on the 4d coordinates through the moduli $v^a(x)$ parameterizing the K\"ahler form \hbox{$J= v^a\om_a$} ($\,\{\om_a(v)\}$ is a basis of harmonic (1,1)--forms): the relation is \hbox{$vol_6 = \textstyle{\frac{1}{6}} J\wedge J\wedge J$}. Therefore one has\footnote{Notice that even if the harmonic forms $\om_a$ depend on the moduli, as illustrated in \cite{MinasianKashani, KashaniPoorNearlyKahler} on a Calabi-Yau one has $v^b \pd{\,}{v^a}\om_b=0$, and therefore $\pd{\,}{v^a} J = \om_a$.}
\be\label{eq:VariationSqrtgCY}
\partial_\mu \log \sqrt{g_6} = \pd{\,}{v^a}(\log \sqrt{g_6}) \partial_\mu v^a = 3 \frac{\om_a \wedge J\wedge J}{J\wedge J\wedge J}\partial_\mu v^a = (\om_a \lrcorner J) \partial_\mu v^a\;.
\ee
The statement then follows recalling that below eq.$\:$(\ref{eq:constancyOfOmContrJ}) we deduced $\partial_m(\om_a \lrcorner J)=0$.

$\cl V_{\mathrm{NS}}$ is zero due to the Ricci-flatness of Calabi-Yaus, as well as to the harmonicity of $\phi$ and $b$. The absence of a scalar potential in the 4d effective action (there is no contribution from the RR sector either) is consistent with the fact that the dimensional reduction is performed on a class of equivalent \emph{solutions} of the 10d theory (with vanishing 4d cosmological constant), so that the geometrical moduli correspond to massless 4d scalars with no preferred vev. This is in contrast with what expected for general \sst structure off-shell reductions: as we will discuss in subsection$\:$\ref{ScalarPotential}, in this case a non-trivial scalar potential is generated.

\subsection{Scalar kinetic terms}\label{ScalarKinTerms}

The second line of (\ref{eq:decompositionNSsector}) defines the kinetic terms for the internal metric and $b$-field fluctuations along the 4d spacetime. This was already translated in the generalized geometry formalism in \cite{CassaniBilal}, where we showed that
\be\label{eq:metricModuliSpace}
\frac{1}{8} g^{mn}g^{pq}(\delta g_{mp}\delta g_{nq} + \delta b_{mp}\delta b_{nq}) \;=\; - \frac{\langle \delta\chi_-, \delta \bar\chi_-\rangle}{\langle \Phi_-, \bar\Phi_-\rangle} - \frac{\langle \delta\chi_+, \delta \bar\chi_+\rangle}{\langle \Phi_+, \bar\Phi_+\rangle}\;.
\ee
In the following we add a comment on this formula. In \cite{CassaniBilal} since the beginning we discarded pure spinor deformations living in the vector representation of O(6,6), decomposing under SU(3)$\times$SU(3) in the `triplets' $({\bf 3},{\bf 1}) \oplus ({\bf {\bar 3}},{\bf 1})\oplus ({\bf 1},{\bf 3}) \oplus ({\bf 1},{\bf{\bar 3}})$. However, eq.$\:$(\ref{eq:metricModuliSpace}) is correct even when taking such variations $\delta_{\mathrm{tr}} \Phi_\pm$ into account, because they are precisely the ones which modify the compatible pair of generalized almost complex structures $\cal J_+,\cal J_-$ while leaving invariant the generalized metric $\cal G= -\cal J_+\cal J_-$ (and therefore the internal metric and $b$--field, see subsection$\:$\ref{SugraFieldsAndsst}). Indeed, recalling the comment below eq.$\:$(\ref{eq:TT*metric}), the space of compatible $\cal J_+,\cal J_-$ at a point of $M_6$ is the 48-dimensional coset $\frac{O(6,6)}{U(3)\times U(3)}$, while the space of generalized metrics $\cal G$ is the 36-dimensional coset $\frac{O(6,6)}{O(6)\times O(6)}$. The 48 -- 36 = 12-dimensional space of transformations being in the first but not in the second coset is in correspondence with the O(6,6) vectors \cite{GLW1}. 

This argument can be made more explicit as follows. Consider the pure spinor variations $\delta_{\mathrm{tr}} \Phi_\pm$ in the SU(3)$\times$SU(3) `triplets', parameterized as in (\ref{eq:DeformationsInTriplets}). Starting from (\ref{eq:VarJ}), we now evaluate the corresponding deformations of the generalized almost complex structures $\cal J_+$ and $\cal J_-$. Performing the computation in the bispinor picture via the same procedure used in \cite{CassaniBilal} to derive eq.$\:$(\ref{eq:metricModuliSpace}) here above, we find that:
\be
-(\delta_{\mathrm{tr}} {\cal J}_+){\cal J}_-  \,=\, \cl B \left(\begin{array}{cc} \im( \delta u\lrcorner\Om_1 + \delta v\lrcorner \Om_2 )^{m}_{\;\;\,n}  & \im( \delta u\lrcorner\Om_1 - \delta v\lrcorner \Om_2 )^{mn} \\ [2mm] \im( \delta u\lrcorner\Om_1 - \delta v\lrcorner \Om_2 )_{mn} &  \im( \delta u\lrcorner\Om_1 + \delta v\lrcorner \Om_2 )_m^{\;\;\,n} \end{array}\right){\cl B}^{-1}  \,=\, + \cal J_+ (\delta_{\mathrm{tr}} \cl J_-)\,,
\ee
where $\Om_1$ and $\Om_2$ are the invariant $(3,0)$--forms for the SU(3) structures associated with $\eta^1_+$ and $\eta^2_+$ respectively (see subsect.$\:$\ref{SU3strConv} of the appendix for our conventions). Therefore we conclude that $\cal G = -\cal J_+\cal J_-$ is invariant under $\delta_{\mathrm{tr}} \Phi_\pm$.

As discussed in subsection$\:$\ref{DefiningTruncation}, the requirement of dropping the pure spinor deformations $\delta_{\mathrm{tr}} \Phi_\pm$ makes (\ref{eq:metricModuliSpace}) coincide with the sum of two special K\"ahler metrics.

Recalling (\ref{eq:PureSpNorm}) and the definition of the 4d dilaton (\ref{eq:4dDilaton}), we can also integrate (\ref{eq:metricModuliSpace}) over the compact $M_6$, and write
\be\label{eq:MetricgAndbDeform}
\frac{e^{2\varphi}}{8} \int vol_6 e^{-2\phi} g^{mp}g^{nq}\big( \delta g_{mn}\delta  g_{pq} + \delta  b_{mn}\delta  b_{pq} \big)= - \frac{\int\langle \delta\chi_-, \delta \bar\chi_-\rangle}{\int\langle \Phi_-, \bar\Phi_-\rangle} - \frac{\int\langle \delta\chi_+, \delta \bar\chi_+\rangle}{\int\langle \Phi_+, \bar\Phi_+\rangle}\;.
\ee
In \cite{CassaniBilal} we parameterized $\delta\chi_\pm$ in terms of the finite set of modes surviving the truncation as $\delta\chi_- = \chi^-_i\delta z^i$ and $\delta\chi_+ = \chi^+_a\delta t^a$, where $z^i$ and $t^a$ are special coordinates for $\mathscr M_-$ and $\mathscr M_+$ respectively. We then concluded that the second line of (\ref{eq:decompositionNSsector}) can be rewritten as the sum $g^-_{i\bar \jmath}\partial_\mu z^i\partial^\mu \bar z^{\bar \jmath} \,+\, g^+_{a\bar b}\partial_\mu t^a\partial^\mu \bar t^{\bar b}$, involving the special K\"ahler metrics $g^-_{i\bar \jmath}$ and $g^+_{a\bar b}$ obtained deriving the K\"ahler potentials (\ref{eq:Kpm}).

\subsection{Variations of $\sqrt{g_6}$ and the dilaton}

In this subsection we discuss the condition under which the variation of $\log(e^{-2\phi} \sqrt{g_6})$, as induced by \sst structure deformations, is independent of the internal coordinates. As observed above eq.$\:$(\ref{eq:VariationSqrtgCY}), this guarantees vanishing of the third line in (\ref{eq:decompositionNSsector}), in analogy with the Calabi-Yau case. 

Recalling the stated relation (\ref{eq:PureSpNorm}) between the dilaton $\phi$ and the pure spinor norm, we immediately see that under a general pure spinor deformation (\ref{eq:variationPhi}) we have
\be
\delta\log(e^{-2\phi} \sqrt{g_6}) = \frac{\delta \langle \Phi_\pm, \bar\Phi_\pm\rangle}{\langle \Phi_\pm, \bar\Phi_\pm\rangle} = 2\re(\delta\kappa)\;,
\ee
where we call $\re(\delta\kappa)$ the equal real parts of $\delta\kappa_+$ and $\delta\kappa_-$. Thus we need constantness along $M_6$ of the function $\re(\delta\kappa)$ associated with pure spinor rescalings. For the truncated set of modes, this is guaranteed by our assumption (\ref{eq:KappaConstant}).

Notice from (\ref{eq:PureSpNorm}) that a priori the metric deformations also affect the dilaton, in such a way that $e^{-2\phi} \sqrt{g_6}$ is left invariant. However, it is more natural to consider the deformations of $\phi$ and $\sqrt{g_6}$ as independent. This can be achieved as follows. We start deriving the first order variation of $\sqrt{g_6}$ induced by $\delta\Phi_\pm$ in (\ref{eq:variationPhi}). Recalling (\ref{eq:TT*metric}), and assuming here $b=0$ for simplicity, we have that $g_{mn} = \cl G_{mn} = -(\cl J_+\cl J_-)_{mn}$. Using (\ref{eq:CurlyJ+-}) we obtain\footnote{If $b=0$ in $\cl G$, then in general the variation $\delta \cl G $ will contain a small $\delta b$. However, at first order this doesn't enter in $\delta\cl G_{mn}$, which is then identified with $\delta g_{mn}$.}$^,$\footnote{The supplementary term $g^{mn}[(\delta \cl J_+)_m^{\;\;p}\cl J_{-pn} + \cl J_{+mp}(\delta \cl J_-)^{p}_{\;\,n}]$ that should enter in (\ref{eq:LinkDeltaSqrtgDeltaCalJ}) vanishes because $g^{nm}(\delta \cl J_+ -  \delta_{\mathrm{tr}}\cl J_+)_m^{\;\;p}$ and $(\delta \cl J_- -  \delta_{\mathrm{tr}}\cl J_-)^{p}_{\;\,n}g^{nm}$ turn out to be symmetric tensors while $\cl J_{-pn}$ and $\cl J_{+mp}$ are antisymmetric.}
\be\label{eq:LinkDeltaSqrtgDeltaCalJ}
2\,\delta \log\sqrt{g_6} \equiv g^{mn}\delta g_{mn} = \frac{1}{2}\big[ (\delta\cl J_+)_{mn}(J_1+J_2)^{mn} + (\delta\cl J_-)_{mn}(J_1-J_2)^{mn} \big]\;,
\ee
so we see that in general both $\delta\cl J_+$ and $\delta\cl J_-$ will contribute. Now we express $\delta \cl J_\pm$ employing (\ref{eq:VarJ}): as discussed in the previous subsection, $\delta_{\mathrm{tr}}\cl J_\pm$ drop when computing variations of the generalized metric $\cl G$, so we are left with the deformations induced by $\delta \chi_\pm$. Performing as usual the computation in the bispinor picture and recalling (\ref{eq:projectorsDeltaChi+-}) we arrive at the result:
\be
\delta \log\sqrt{g_6} = 4g^{mn}\re  (\delta \chi_- - \delta \chi_+)_{mn}\;.
\ee
Recalling (\ref{eq:PureSpNorm}), we can now prevent such a metric variation to modify the dilaton $\phi$ by prescribing a simultaneous real rescaling of $\Phi_\pm$ with $\delta\kappa =  \textstyle{\frac{1}{2}}\delta \log \sqrt {g_6}$. Any other independent pure spinor rescaling (having $\re(\delta\kappa)\neq 0$) modifies $\phi$ without affecting the metric $g_{mn}$.

All this can be illustrated considering strictly SU(3) structures. In this case $J_1=J_2\equiv J$ and $I_1=I_2\equiv I$, so that from (\ref{eq:CurlyJ+-}) we have $\cl J_+ = { {0\;\: -J^{-1}} \choose {\!\!\!J\;\;\;\;\; 0} }$ and $\cl J_- = { {\!\!-I\;\; 0} \choose {\:0\;\; \,I^T} }$. From (\ref{eq:LinkDeltaSqrtgDeltaCalJ}) we immediately see that $\delta \cl J_-$ does not contribute, and that
\be
\,\delta \log\sqrt{g_6} = \frac{1}{2}(\delta\cl J_+)_{mn}  J^{mn} = (\delta J)\lrcorner J\;.
\ee
In particular, only the rescalings $\delta J=\delta\lambda J$ (where $\delta\lambda$ is a function) contribute to $(\delta J)\lrcorner J$. Now we notice that this $J$--rescaling also implies a rescaling of $\Phi^0_+$, which in the SU(3) structure case reads $\Phi^0_+= e^{-\phi}e^{-iJ}$ (recall (\ref{eq:SU3PureSpinors}) and (\ref{eq:ProductNormsIsDilaton})). Indeed, at first order we have
\be\label{eq:DeltaPhiUnderDeltaJ}
\delta e^{-iJ} = \frac{3}{2} \delta \lambda e^{-iJ} + \frac{1}{4}\delta\lambda (-6 + 2iJ - J^2 + iJ^3)\;,
\ee
where the second term in the rhs is in the $(\bf{\bar 3}, \bf{3})$ of SU(3)$\times$ SU(3). It is now immediate to check that, thanks to the presence of the rescaling term in (\ref{eq:DeltaPhiUnderDeltaJ}), it is consistent to keep the pure spinor norm (\ref{eq:PureSpNorm}), viz. the dilaton, unmodified. Thus the condition $\re(\delta\kappa)=\mathrm{const}$ in this case also requires $\delta\lambda$ to be constant along $M_6$. Choosing the basis of expansion forms described in \cite{MinasianKashani}, we have $3\delta\lambda=(\delta J)\lrcorner J = \om_a\lrcorner J\delta v^a$, and we recover the requirement $d(\om_a\lrcorner J)=0$ discussed in that paper (as seen below (\ref{eq:constancyOfOmContrJ}), this is satisfied for a Calabi-Yau).

\subsection{Scalar potential}\label{ScalarPotential}
In the following first we obtain a formula expressing the Ricci curvature $R_6$ of the compact manifold (supplemented by terms involving $H_{mnp}$ and $\partial_m\phi$) as a function of the pure spinors $\Phi_\pm$. Then we apply this result to reformulate the NSNS contribution (\ref{eq:DefV_NS}) to the 4d scalar potential. This allows us to make contact with an expression for the potential obtained with purely 4d gauged supergravity methods in \cite{D'AuriaFerrTrigiante}.

At the end of this subsection we will prove that under the assumption
\be\label{eq:NoSomeTripletsIndPhi}
\langle d_H \Phi^0_+ , \buildrel\to\over\gamma\!{}^{m} \bar \Phi^0_+ \rangle + \langle d_H \Phi^0_- , \buildrel\to\over\gamma\!{}^{m} \bar \Phi^0_- \rangle  =  0 \quad ,\quad  \langle d_H \Phi^0_+ ,  \bar \Phi^0_+\!  \buildrel\leftarrow\over\gamma\!{}^{m}\rangle + \langle d_H \bar \Phi^0_- , \Phi^0_- \! \buildrel\leftarrow\over\gamma\!{}^{m} \rangle  =  0\;,
\ee
constraining a subset\footnote{Here we don't strictly need the condition projecting out \emph{all} the \sst triplets in $d_H \Phi^0_\pm$, which would read: $\langle d_H \Phi^0_+ , \Gamma^\Lambda \bar \Phi^0_+ \rangle\; = \; 0 \; =\; \langle d_H \Phi^0_- , \Gamma^\Lambda \bar \Phi^0_- \rangle\,,\;\, \textrm{with }\Gamma^{\Lambda}= dy^m\!\!\wedge\;\textrm{or}\; \iota_{\partial_m}$ (the analogous relations containing $\Phi^0_\pm$ at the place of $\bar\Phi^0_\pm$ are automatically satisfied).} of the SU(3)$\times$SU(3) triplets in $d_H\Phi^0_\pm$, the following formula is valid:
\bea\label{eq:FormulaForInternalNSsector}
&& R_6 -\frac{1}{12} H_{mnp}H^{mnp}  + 4 \partial_m\phi \partial^m\phi  - 2e^{2\phi} \nabla_{\!6}^2 e^{-2\phi} =\\ [2mm]
\nnb &=&\!\!\! -4 \frac{\langle d_H \Phi^0_+, *\lambda(d_H \bar\Phi^0_+) \rangle}{i \langle \Phi_\pm,\bar\Phi_\pm\rangle} -4 \frac{\langle d_H \Phi^0_- , *\lambda(d_H \bar\Phi^0_-) \rangle}{i \langle \Phi_\pm,\bar\Phi_\pm\rangle} +  16\Big|\frac{\langle d_H\Phi^0_+ , \Phi^0_-\rangle}{i \langle \Phi_\pm,\bar\Phi_\pm\rangle}\Big|^2   + 16\Big|\frac{\langle d_H \Phi^0_+ , \bar\Phi^0_-\rangle}{i \langle \Phi_\pm,\bar\Phi_\pm\rangle}\Big|^2,
\eea
where $ \nabla_{\!6}^2$ is the laplacian on $M_6$ and $d_H= d-H\wedge$, with $H=H^{\mathrm{fl}}+ d_{(6)}b$ purely internal. This completes and generalizes an expression given in the context of SU(3) structures in footnote$\:$2 of ref.$\:$\cite{MinasianKashani}, referring to results in \cite{BedulliVezzoni}.

We remark that (\ref{eq:FormulaForInternalNSsector}) is symmetric under the exchange $\Phi^0_+\leftrightarrow \Phi^0_-$, in agreement with the formulation of mirror symmetry in the context of generalized structures \cite{FidanzaMinasianTomasiello, GMPT1, GMPT2}. Indeed we have $\langle d_H \Phi^0_+ , \Phi^0_-\rangle = \langle \Phi^0_+, d_H \Phi^0_- \rangle$, thanks to the fact that $\Phi^0_+,\Phi^0_-$ satisfy (\ref{eq:compatibility}). 

Furthermore, notice that while the last two terms in the rhs of (\ref{eq:FormulaForInternalNSsector}) are positive definite, the first two are instead negative definite: in fact for any complex polyform $C= \sum_k C_k$, one has $\langle C,  *\lambda(\bar C)\rangle = vol_6\sum_k C_k \lrcorner \bar C_k$. The last two terms of (\ref{eq:FormulaForInternalNSsector}) vanish when at least one of the two pure spinors satisfies the integrability condition $d_H\Phi^0 = (\iota_v + \zeta\wedge)\Phi^0$, where $v$ is a vector and $\zeta$ a 1--form. Finally, the rhs of (\ref{eq:FormulaForInternalNSsector}) vanishes identically when the pure spinors satisfy the `generalized Calabi-Yau metric' condition $d_{H}\Phi^0_\pm = 0$ introduced in \cite{GualtieriThesis}.\footnote{This condition does not coincide with the notion of generalized Calabi-Yau manifold defined in \cite{HitchinGenCY}, see e.g. \cite{LiDeformGenCY} (sect.$\:$4) for a comparison.} Then for these geometries we have an expression for the curvature $R_6$ in terms of $\partial_m\phi$ and $H_{mnp}$ (playing the role of torsion).

\vskip .2cm

The rhs of (\ref{eq:FormulaForInternalNSsector}) can also be expressed in terms of the \sst torsion classes introduced in \cite{GMPT2, GMPT3}. We refer to the parameterization provided by eqs.$\:$(6.14), (6.15) of ref.$\:$\cite{GMPT3} (even if written for SU(3) structure pure spinors, that parameterization also applies to the general \sst structure case). Using (\ref{eq:*lambda}), (\ref{eq:MukaiUnderClifford}) we get:
\bea
\textrm{rhs of (\ref{eq:FormulaForInternalNSsector}) } &=& \\ [1mm]
\nnb = \;8\big(\,|W^{30}|^2 \!\!\! &+&\!\!\! |W^{03}|^2 \,\big) - 16\big(\, |W^{21}|^2 + |W^{12}|^2 + |W^{11}|^2 + |W^{22}|^2 + |W^{10}|^2 + |W^{01}|^2\, \big)\;,
\eea
where expressions like for instance $|W^{12}|^2$ and $|W^{10}|^2$ mean $W^{12}_{i_1j_2}\overline{W}^{12\,i_1j_2}$ and $W^{10}_{j_2} \overline{W}^{10\,j_2}$ respectively. As in subsection$\:$\ref{sstDeformations}, the indices $\bar\imath_1, i_1$ are (anti)holomorphic with respect to the almost complex structure $I_1$, and analogously for $\bar\jmath_2, j_2$ w.r.t. $I_2$. Our constraint (\ref{eq:NoSomeTripletsIndPhi}), which in terms of torsion classes reads $W^{01}_{\bar\imath_1} + W^{31}_{\bar\imath_1}=0$ and $W^{10}_{j_2} - \overline{W}^{20}_{j_2}=0$, has been used to eliminate $W^{31}$ and $W^{20}$.

\vskip .2cm
Now we multiply eq.$\:$(\ref{eq:FormulaForInternalNSsector}) with $e^{-2\phi} vol_6\, $ and integrate over $M_6$, getting in this way a geometric expression for the NSNS contribution (\ref{eq:DefV_NS}) to the 4d scalar potential:
\begin{eqnarray}
\nnb \cl V_{\mathrm{NS}} &=&  \;\frac{e^{4\varphi}}{4}\int \Big[\, \langle \,d_{H^{\mathrm{fl}}} \Phi_+ , *_b (d_{H^{\mathrm{fl}}} \bar\Phi_+) \,\rangle +  \langle\, d_{H^{\mathrm{fl}}} \Phi_- , *_b (d_{H^{\mathrm{fl}}} \bar\Phi_-) \,\rangle\, \Big]\\ [2mm] 
\label{eq:V_NSpureSpinors}&& -\; e^{4\varphi}\int \frac{\big |\langle d_{H^{\mathrm{fl}}} \Phi_+ , \Phi_-\rangle\big |^2  +  \big |\langle d_{H^{\mathrm{fl}}} \Phi_+ , \bar\Phi_-\rangle\big |^2 }{i \langle \Phi_\pm,\bar\Phi_\pm \rangle} \;\;.
\end{eqnarray}
Eq.$\:$(\ref{eq:PureSpNorm}) has been used, as well as (\ref{eq:bTransform}) and the definition (\ref{eq:Def*b}) of the $*_b$ operator.

Starting from (\ref{eq:V_NSpureSpinors}), it is possible to reformulate $\cl V_{\mathrm{NS}}$ in terms of the 4d degrees of freedom by substituting the expansions (\ref{eq:expansPhi+-}) for $\Phi_\pm$ and exploiting the assumed properties of the basis polyforms. For instance, recalling (\ref{eq:AnsatzDerForms}), (\ref{eq:mathbbMandmathbbN}):
\be
e^{2\varphi}\int  \langle d_{H^{\mathrm{fl}}} \Phi_+ , *_b (d_{H^{\mathrm{fl}}} \bar\Phi_+) \rangle = - 8e^{K_+} X^{\mathbb A}(\mathbb{Q}^T \widetilde{\mathbb M} \mathbb{Q})_{\mathbb{AB}}\bar X^{\mathbb B}\;,
\ee
where we have also used the fact that $e^{-K_\pm}=  8e^{-2\varphi}$ (see (\ref{eq:PureSpNorm}), (\ref{eq:Kpm}) and (\ref{eq:4dDilaton})). To evaluate the second line of (\ref{eq:V_NSpureSpinors}), we need requirement (\ref{eq:RatiosConstant}), implying:
\be
\frac{\langle \Sigma_-^{\mathbb I} , \Phi_- \rangle}{\langle \Phi_-, \bar \Phi_- \rangle} = \frac{\int\langle \Sigma_-^{\mathbb I} , \Phi_- \rangle}{\int\langle \Phi_-, \bar \Phi_- \rangle} = -i e^{K_-}Z^{\mathbb I}\;.
\ee
The resulting expression for $\cl V_{\mathrm{NS}}$ is symplectically invariant, and reads:
\bea
\nnb \cl V_{\mathrm{NS}} &=& - \;2 e^{2\varphi} \Big[ e^{K_+} X^{\mathbb A}(\mathbb{Q}^T \widetilde{\mathbb M} \mathbb{Q})_{\mathbb{AB}}\bar X^{\mathbb B} +  e^{K_-}Z^{\mathbb I}(\widetilde{\mathbb Q}^T \widetilde{\mathbb N} \widetilde{\mathbb Q})_{\mathbb{IJ}}\bar Z^{\mathbb J}  \Big]\\ [1mm]
\label{eq:VNSinTermsOf4dFields}&& -\;8 e^{2\varphi}e^{K_+ + K_-} \bar Z^{\mathbb I}(\mathbb S_-\mathbb Q)_{\mathbb{IA}}(X^{\mathbb A}\bar X^{\mathbb B} +\bar X^{\mathbb A} X^{\mathbb B} )(\mathbb{Q}^T\mathbb S_-)_{\mathbb{BJ}}Z^{\mathbb J}\;,
\eea
where we recall that $\widetilde{\mathbb Q}$ is given by (\ref{eq:ChargeMatrixTildeQ}) and that $\widetilde{\mathbb N}$ and $\widetilde{\mathbb M}$ are negative definite. This is precisely the same expression obtained in \cite{D'AuriaFerrTrigiante} by means of 4d gauged supergravity techniques, starting from the 4d effective action associated with Calabi--Yau compactifications.

\vskip .2cm

Finally, we remark that the value of expression (\ref{eq:FormulaForInternalNSsector}) {\it in a vacuum} is also related to the external spacetime Ricci curvature $R_4$. Indeed the 10d dilaton equation (in string frame and in the absence of localized sources) for a 4d$\times$6d background preserving maximal 4d symmetry takes the form
\be\label{eq:DilatonEqOnMaxSymm4d}
-R_4 = R_6 -\frac{1}{12} H_{mnp}H^{mnp}  + 4 \partial_m\phi \partial^m\phi  - 2e^{2\phi} \nabla_6^{\:2}e^{-2\phi}\;,
\ee
with no contributions from the RR sector. Furthermore, acting on eq.$\:$(\ref{eq:DilatonEqOnMaxSymm4d}) with $\int_{M_6}e^{-2\phi} vol_6 $ and rescaling the 4d metric as in (\ref{eq:rescale4dmetric}), we obtain $R_4 = 2\cl V_{\mathrm{NS}}$. On the other hand, from the trace of the 4d Einstein equation evaluated on a maximally symmetric vacuum, in general one has $R_4 = 4\cl V$. Since the total potential of the reduced theory is $\cl V= \cl V_{\mathrm{NS}} + \cl V_{\mathrm{RR}}$, then we can conclude that in a vacuum $2\cl V_{\mathrm{RR}} = - \cl V_{\mathrm{NS}}$.

\vskip .4cm

\underline{\emph{Proof of relation (\ref{eq:FormulaForInternalNSsector})}}\\ [-2mm]

In the remainder of this section we give an account of the main computational steps proving eq.$\:$(\ref{eq:FormulaForInternalNSsector}). We parameterize $||\eta_\pm^1|| = |a|,\;||\eta_\pm^2|| = |b|$ (this last should not be confused with the internal NS 2--form, also called $b$). Then (\ref{eq:ProductNormsIsDilaton}) says $|ab|= e^{-\phi}$.

We start without imposing any constraint on the SU(3)$\times$SU(3) triplets of $d_H{\Phi_\pm^0}$. The rhs of (\ref{eq:FormulaForInternalNSsector}) is evaluated using (\ref{eq:MukaiUnderClifford}), (\ref{eq:PSnorm}) for the Mukai pairing as well as 
\bea\label{eq:slashdHPhi}
\nnb \frac{1}{4}
\rlap{\begin{picture}(10,10)(-2,1)
\put(0,0){\line(2,1){20}}
\end{picture}}
d_H\Phi^0_{\pm} &=& (\slashD - {\textstyle\frac{1}{4}}\slashH) \eta_+^1\eta_\pm^{2\dag}\, \pm \,(D_m- {\textstyle\frac{1}{4}} H_m)\eta_+^1\eta_\pm^{2\dag}\gamma^m\\ 
&\pm& \eta_+^1 \big[ (\slashD + {\textstyle\frac{1}{4}}\slashH) \eta_\pm^2 \big]^\dag \, + \, \gamma^m\eta_+^1\big[ (D_m + {\textstyle\frac{1}{4}} H_m)\eta_\pm^2 \big]^\dag\;,
\eea
where $\slashD=\gamma^nD_n\,,\;\slashH = \frac{1}{6}H_{mnp}\gamma^{mnp}$ and $H_m = \frac{1}{2}H_{mnp}\gamma^{np}$. Eq.$\:$(\ref{eq:slashdHPhi}) is directly derived (also recalling (\ref{eq:defPhi0})) from the expressions for $\rlap{\begin{picture}(10,10)(-2,1)
\put(0,0){\line(1,1){10}}
\end{picture}}
d\Phi^0_{\pm}$ and $\!\!\!\!\begin{picture}(10,11)(-15,1)
\put(0,0){\line(3,1){30}}
\end{picture} H\wedge \Phi^0_\pm$ given e.g. in appendix A of ref.$\:$\cite{GMPT3}. For instance we obtain
\bea
\nnb 16\Big|\frac{\langle d_H\Phi^0_+ , \Phi^0_-\rangle}{i \langle \Phi_\pm,\bar\Phi_\pm\rangle}\Big|^2 &=& 4|a|^{-4}\big[(\slashD - {\textstyle\frac{1}{4}}\slashH) \eta_+^1\big]^\dag \eta_-^1\eta_-^{1\dag} (\slashD - {\textstyle\frac{1}{4}}\slashH) \eta_+^1\\ 
&=& 4|a|^{-2} \big|(\slashD - {\textstyle\frac{1}{4}}\slashH) \eta_+^1\big|^2 -2|a|^{-4} \big|\eta_+^{1\dag} \gamma^m (\slashD - {\textstyle\frac{1}{4}}\slashH) \eta_+^1\big|^2\;,
\eea
where for the second equality we used identity (\ref{eq:ChiralityAsProjector}) in order to reexpress $|a|^{-2}\eta_-^1\eta_-^{1\dag}$. The computation of the terms in the rhs of (\ref{eq:FormulaForInternalNSsector}) containing $*\lambda$ is slightly more involved, but employs the same technique. For the image of $*\lambda$ under the Clifford map we use (\ref{eq:*lambda}). 

Resumming all the terms and taking a few cancellations into account we obtain:
\bea
\nnb &-&\!\!\!\!4 \frac{\langle d_H \Phi^0_+, *\lambda(d_H \bar\Phi^0_+) \rangle}{i \langle \Phi_\pm,\bar\Phi_\pm\rangle} -4 \frac{\langle d_H \Phi^0_- , *\lambda(d_H \bar\Phi^0_-) \rangle}{i \langle \Phi_\pm,\bar\Phi_\pm\rangle} +  16\Big|\frac{\langle d_H\Phi^0_+ , \Phi^0_-\rangle}{i \langle \Phi_\pm,\bar\Phi_\pm\rangle}\Big|^2   + 16\Big|\frac{\langle d_H \Phi^0_+ , \bar\Phi^0_-\rangle}{i \langle \Phi_\pm,\bar\Phi_\pm\rangle}\Big|^2\; =\\ [2mm]
\nnb &=& |a|^{-2} \big[ 2 D_m\eta_+^{1\dag}\gamma^{mn}D_n\eta^1_+  + \frac{1}{8}\, \eta_+^{1\dag}( H_mH^m - \slashH \slashH ) \eta_+^1  -  \frac{1}{12} D_m (\eta_+^{1\dag} \gamma^{mnpq}\eta_+^1) H_{npq} \big] \\ [2mm]
\nnb &-&  4|a|^{-2}\re\big[ \eta_+^{1\dag}\gamma^m(\slashD - {\textstyle\frac{1}{4}}\slashH)\eta_+^1  \big] \partial_m\log |b| \,-\, 2|a|^{-4} \big| \eta_+^{1\dag}\gamma^m(\slashD - {\textstyle\frac{1}{4}}\slashH)\eta_+^1 \big|^2 \\ [2mm]
\label{eq:RelWithNoConstraints} &+& \; \eta^1_+ \rightarrow \eta^2_+ \,\;,\;\, |a| \leftrightarrow |b| \;\,,\,\; H\rightarrow -H
\eea
where the last line denotes the repetition of the two preceding lines performing the prescribed transformations.

Now we consider our requirement (\ref{eq:NoSomeTripletsIndPhi}) on the SU(3)$\times$SU(3) triplets of $d_H{\Phi_\pm^0}$: this can be translated as:\footnote{One can check that in the notation of ref.$\:$\cite{GMPT3} (sect.$\:$A.4), this constraint corresponds to $T^1_{\bar\imath_1} + \partial_{\bar\imath_1}\log|b|=0$ together with $T^2_{\bar\imath_2} + \partial_{\bar\imath_2}\log|a|=0$.}
\bea
\label{eq:NoTripletsIndPhiSecond}  |a|^{-2}\eta_+^{1\dag}\gamma^m(\slashD - {\textstyle\frac{1}{4}}\slashH)\eta_+^1 + 2P^m_{1\;n}\partial^n\log |b| &=& 0\;,
\eea
together with the analogous relation obtained implementing $1\rightarrow 2\:,\; |a| \leftrightarrow |b| \:,\; H\rightarrow -H$. $P_1$ is the holomorphic projector associated with the almost complex structure $I_1$.

Now, constraint (\ref{eq:NoTripletsIndPhiSecond}) implies that the two terms in (\ref{eq:RelWithNoConstraints}) containing $\slashD - {\textstyle\frac{1}{4}}\slashH$ cancel each other. Then using the following relations:
\bea\label{eq:RelationsToImplement}
\nnb [D_m,D_n]\eta_+ = \frac{1}{4} R_{mnpq}\gamma^{pq}\eta_+ \quad &\Rightarrow &\quad D_m\eta_+^{\dag}\gamma^{mn}D_n\eta_+ = D_m (\eta_+^{\dag}\gamma^{mn}D_n\eta_+) + \frac{1}{4}||\eta_+||^2 R_6 \\
\nnb H_mH^m - \slashH \slashH  &=& - \frac{1}{3} H_{mnp}H^{mnp}\\
dH=0 \quad & \Leftrightarrow &\quad D_{[m}H_{npq]} = 0\;,
\eea
we rewrite:
\bea
\nnb \label{eq:AlmostResult} \textrm{rhs of (\ref{eq:RelWithNoConstraints})}\! &=& \! R_6 -  \frac{1}{12} H_{mnp}H^{mnp}\: +\: 2|a|^{-2} D_m \big(\eta_+^{1\dag}\gamma^{mn}D_n\eta^1_+ -  \frac{1}{24} H_{npq} \eta_+^{1\dag} \gamma^{mnpq}\eta_+^1 \big)\\ 
&&+\;\;  \eta^1_+ \rightarrow \eta^2_+ \;\,,\,\; |a|\rightarrow |b|  \;\,,\,\; H\rightarrow -H \;,
\eea
where only the term involving $|a|^{-2}$ needs to be repeated with the prescribed substitutions.

Now we observe that the real part of constraint (\ref{eq:NoTripletsIndPhiSecond}) can be written as:
\be
|a|^{-2} \big[\re(\eta_+^{1\dag}\gamma^{mn}D_n \eta_+^1) -\frac{1}{24} H_{npq}\eta_+^{1\dag} \gamma^{mnpq} \eta_+^1 \big] + \partial^m\log |ab| = 0\;.
\ee
Noticing that $D_m[\im(\eta_+^{\dag}\gamma^{mn}D_n\eta_+)]$ vanishes identically, and recalling $|ab| = e^{-\phi}$, we can use this equation, together with the analogous one obtained performing $1\rightarrow 2\:,\; |a| \leftrightarrow |b| \:,\; H\rightarrow -H$, to see that 
\be
\textrm{last two terms in (\ref{eq:AlmostResult})}\;\, =\, -4 \partial_m\phi \partial^m\phi  + 4 \nabla_6^{\:2} \phi\, \equiv\, 4 \partial_m\phi \partial^m\phi  - 2e^{2\phi}  \nabla_6^{\:2} e^{-2\phi}\;.
\ee 
This proves eq.$\:$(\ref{eq:FormulaForInternalNSsector}).

\section{Reduction of the RR sector}\label{ReductionRRsector}

\setcounter{equation}{0}

In this section we reduce the RR sector. We will focus on type IIA, but the procedure we describe can equally well be applied to type IIB. 

We wish to reduce the RR democratic pseudo-action (\ref{eq:RRpseudoAction}), also implementing the self-duality constraint (\ref{eq:10dSelfDualityF}) in an appropriate way (a direct substitution of (\ref{eq:10dSelfDualityF}) in (\ref{eq:RRpseudoAction}) results indeed in a vanishing action). In principle we could follow a procedure similar to the one adopted in \cite{DallAgataHfluxes}, and subsequently in \cite{LouisMicu, GurrMicuIIBhalfFlat}, to reduce the type IIB action taking into account the self-duality of the RR 5--form $F_5$. In \cite{DallAgataHfluxes}, first the electric and magnetic 4d gauge field strengths descending from the expansion of $F_5$ on the Calabi-Yau harmonic 3--forms are regarded as independent and kept in the 4d action. Then the addition of a suitable Lagrange multiplier term makes the equations of motion for the magnetic field strengths precisely correspond to the self-duality constraint. Integrating out the magnetic field strengths provides thus an action with electric fields only and the self-duality constraints correctly implemented. In our context, the generalization of this procedure would require to keep in the 4d action forms of any degree\footnote{It would be interesting to relate this with the tensor hierarchy proposed in \cite{DeWitNicolaiSamtleben}.} (from 0 to 4) descending from the RR field expansion on the internal basis (\ref{eq:Sigma+-AsVectors}), and then to integrate out a subset of these forms.

However in our case this direct approach to the reduction of the action turns out to be quite involved due to the large amount of fields and constraints, and indeed we find it more efficient to proceed along the following alternative path. 

First we reduce the self-duality constraint for the democratic RR field, as well as its EoM/Bianchi identities. From the reduced Bianchi identities we isolate and solve a set of 4d Bianchi identities, defining in this way the fundamental dynamical fields of the 4d effective theory. Using the relations obtained from the reduction of the 10d self-duality condition, the remaining 4d equations are interpreted as EoM associated with the identified dynamical degrees of freedom. The last step consists in the reconstruction of the four dimensional action leading precisely to such EoM.

We will work with the so-called G--basis for the RR field, defined via \cite{Democratics}:
\be\label{eq:GbasisRRfield}
{\bf \hat F} \equiv e^{\hat B}{\bf \hat G} \;.
\ee
In this basis, the self-duality constraints (\ref{eq:10dSelfDualityF}) and the Bianchi identities in (\ref{eq:10dDemocraticRReom/Bianchi}) read respectively:
\be\label{eq:selfDualityG}
e^{\hat B}{\bf \hat G }=  \lambda * (e^{\hat B}\hat{\bf G})\;,
\ee
\be\label{eq:BianchiForDemocrRRFieldStr}
(d - H^{\mathrm{fl}}\wedge) {\bf \hat G} = 0\;,
\ee
where as in the previous section we used the decomposition $\hat H =  H^{\mathrm{fl}} + d\hat B$, with $\hat B= B + b$. We recall that, due to the self-duality, the RR EoM are equivalent to the Bianchi identities.

\subsection{Reduction of the RR self-duality constraint}\label{RedRRselfDuality}

We start expanding the RR field ${\bf \hat{G}}$ on the internal basis polyforms (\ref{eq:Sigma+-AsVectors}). Recalling (\ref{eq:DemocraticRRfield}) and (\ref{eq:GbasisRRfield}), this expansion naturally leads to forms of any degree in the 4d spacetime $M_4$:
\be\label{eq:ExpansionWithG}
2^{-1/2}{\bf \hat{G}} = (G_0^A + G_2^A + G_4^A) \om_A - (\tilde G_{0A} + \tilde G_{2A} + \tilde  G_{4A}) \tilde \om^A + (G_1^I + G_3^I)\alpha_I - (\tilde G_{1I} + \tilde G_{3I})\beta^I,
\ee
where $G_p$ denotes a $p$--form on $M_4$ depending on the $x^{\mu}$ coordinates only. The $2^{-1/2}$ factor is introduced just for later convenience (concerning the relative normalization of the reduced RR and NSNS sectors). We also introduce the following auxiliary expansion:
\be\label{eq:AuxiliaryExp}
2^{-1/2} e^{\hat B}{\bf \hat{G}}= e^{b}\big( K^A\om_A -\tilde K_A\tilde\om^A + L^I\alpha_I - \tilde L_I \beta^I \big)\,,
\ee
so that (the indices are understood and $B$ is along $M_4$): $\:L = G_1 + (G_3 + B\wedge G_1)$\,, $K = G_0 + (G_2 + B G_0) + (G_4 + B\wedge G_2 + {\textstyle\frac{1}{2}}B\wedge B G_0)$, and analogously for $\tilde K$ and $\tilde L$.
\vskip .3cm
We now reduce the self-duality constraint (\ref{eq:selfDualityG}). Substituting (\ref{eq:AuxiliaryExp}), this can be rewritten as:
\be\label{eq:SelfDualityAfterExp}
K^A\om_A -\tilde K_A\tilde\om^A + L^I\alpha_I - \tilde L_I \beta^I =  -*\lambda(K^A)*_b\om_A +*\lambda(\tilde K_A)*_b\tilde\om^A - *\lambda(L^I)*_b\alpha_I + *\lambda(\tilde L_I) *_b\beta^I
\ee
where (\ref{eq:*lambdaSplits}) has been used, as well as the definition (\ref{eq:Def*b}) of the 6d operator $*_b\,$. Taking the Mukai pairings with the basis forms, integrating over $M_6$ and using the results for the action of $*_b$ recalled in subsection$\:$\ref{DefiningTruncation}, from (\ref{eq:SelfDualityAfterExp}) we get the 4d relations:
\begin{eqnarray}
\label{eq:selfdualityOnQ}\tilde K_A &=& -\im\mathcal N_{AB} *\lambda(K^B) + \re\mathcal N_{AB} K^B\\ [1mm]
\label{eq:sdualityOnR}\tilde L_I &=& -\im\mathcal M_{IJ} *\lambda(L^J) + \re\mathcal M_{IJ} L^J\;.
\end{eqnarray}
In order to keep the notation of the forthcoming expressions as compact as possible, we use the symplectic notation introduced in subsection$\:$\ref{DefiningTruncation}, and we define the symplectic vectors 
\be 
G_k^{\mathbb A} = {{G_k^A} \choose {\tilde G_{kA}}}\quad \textrm{for}\; k= 0,2,4 \qquad\textrm{and}\qquad G_k^{\mathbb I} = {{G_k^I} \choose {\tilde G_{kI}}} \quad\textrm{for}\; k= 1,3\;.\ee
Then separating the different form degrees and rescaling the 4d metric as done in (\ref{eq:rescale4dmetric}) for the NSNS sector, (\ref{eq:selfdualityOnQ}) yields the following relations among the 4d fields:
\bea
\label{eq:constraint2-2}\tilde G_{2A} + B \tilde G_{0A}\; =\; \im \mathcal N_{AB} *(G_2^B + BG_0^B)\!\! &+&\!\! \re\mathcal N_{AB} (G_2^B+BG_0^B)\;,\;\;\\ [2mm]
\label{eq:constraint4-0} G_4^{\mathbb A} + B\wedge G_2^{\mathbb A} + \frac{1}{2} B\wedge B G_0^{\mathbb A} &=& e^{4\varphi}\mathbb {N}^{\mathbb A}_{\;\;\mathbb B} G_0^{\mathbb B} \,*1\;,
\eea
while from (\ref{eq:sdualityOnR}) we obtain:
\be\label{eq:constr3-1}
G_3^{\mathbb I} + B \wedge G_1^{\mathbb I} = -e^{2\varphi}\mathbb {M}^{\mathbb I}_{\;\,\mathbb J}* G_1 ^{\mathbb J}\;.
\ee
Eqs.$\:$(\ref{eq:constraint2-2})--(\ref{eq:constr3-1}) represent the 4d remains of the 10d RR self-duality condition (\ref{eq:selfDualityG}).

\subsection{Reduction of the equations of motion / Bianchi identities}

We now pass to reduce eq.$\:$(\ref{eq:BianchiForDemocrRRFieldStr}). This will provide a set of Bianchi identities for the 4d fields as well as the 4d EoM, once the relations (\ref{eq:constraint2-2})--(\ref{eq:constr3-1}) imposed by the reduced 10d self-duality will be used to eliminate the redundant 4d fields. Starting from the expansion (\ref{eq:ExpansionWithG}) for ${\bf \hat G}$, we use the ansatz (\ref{eq:AnsatzDerForms}) to evaluate $d_{H^{\mathrm{fl}}}$ on the internal basis of forms,\footnote{Due to their moduli dependence, the basis forms are in general not closed even with respect to the 4d exterior derivative. However, recall that in subsection$\:$\ref{DefiningTruncation} we assumed that their derivatives with respect to the moduli vanish in the integrated Mukai pairing.} and then separate the different components by acting with $\int_{M_6}\langle \Sigma_\pm,\,\cdot \,\rangle\, $. The following set of four-dimensional equations is obtained (recall that $\widetilde{\mathbb{Q}}$ is related to $\mathbb Q$ as in (\ref{eq:ChargeMatrixTildeQ})):
\bea
\label{eq:EoMBianchi0sympl} \mathbb Q^{\mathbb I}_{\;\,\mathbb A} G_0^{\mathbb A}&=& 0\\ [2mm]
\label{eq:EoMBianchi1sympl} d G_0^{\mathbb A} - \widetilde{\mathbb Q}^{\mathbb A}_{\;\;\,\mathbb I} G_1^{\mathbb I} &=&0\\ [2mm]
\label{eq:EoMBianchi2sympl} d G_1^{\mathbb I} + \mathbb Q^{\mathbb I}_{\;\,\mathbb A} G_2^{\mathbb A} &=& 0 \\ [2mm]
\label{eq:EoMBianchi3sympl} d G_2^{\mathbb A} - \widetilde{\mathbb Q}^{\mathbb A}_{\;\;\,\mathbb I} G_3^{\mathbb I} &=&0\\ [2mm]
\label{eq:EoMBianchi4sympl} d G_3^{\mathbb I} + \mathbb Q^{\mathbb I}_{\;\,\mathbb A} G_4^{\mathbb A} &=& 0\;.
\eea
We immediately rewrite eq.$\:$(\ref{eq:EoMBianchi4sympl}): using (\ref{eq:constraint4-0}) and (\ref{eq:constr3-1}) to eliminate $G_4^{\mathbb A}$ and $G_3^{\mathbb I}$, also employing (\ref{eq:EoMBianchi0sympl}), (\ref{eq:EoMBianchi2sympl}) to simplify the expression, we obtain
\be\label{eq:EoMBianchi4symplModifAgain}
- d\big(\, e^{2\varphi}\mathbb {M}^{\mathbb I}_{\;\mathbb{J}}* G_1^{\mathbb J}\, \big) - dB\wedge G_1^{\mathbb I}  + e^{4\varphi}(\mathbb Q \mathbb {N})^{\mathbb I}_{\;\,\mathbb A} G_0^{\mathbb A}\, *1 = 0\;.
\ee
\vskip .2cm

We also need to reduce the ten dimensional EoM (\ref{eq:EoMfor10dimB}) for the NS 2--form $\hat B$, which receives contributions from both the NSNS and the RR sectors. This is an 8--form equation, and we consider just its piece with 2 legs along $M_4$ and six legs along $M_6$. Taking the integral over $M_6$, using the expansions in subsection$\:$\ref{RedRRselfDuality} and recalling (\ref{eq:mathbbMandmathbbN}), (\ref{eq:tildeN}), we arrive at the 4d equation:
\be
\label{eq:4dBEoM} \frac{1}{2} d(e^{-4\varphi}*dB)\, +\, G_0^A\tilde G_{2A}  - \tilde G_{0A} G_2^A +  \tilde G_{1I}\wedge G_1^I = 0\,,
\ee
where the 4d metric has been Weyl rescaled as in (\ref{eq:rescale4dmetric}). This corresponds to the EoM for the 2--form $B$ in the reduced 4d theory.

\subsection{$p_I^A = 0= q^{IA}$ case. SU(3) structure}\label{p=0=qcase}

We pursue the analysis by considering first the simpler case in which {$p_I^A = 0= q^{IA}$, i.e. $\mathbb Q^{\mathbb IA}=0$ (recall (\ref{eq:ChargeMatrixQ})). As we will discuss below, this is particularly relevant for dimensional reductions on SU(3) structure manifolds.

We start by identifying and solving a set of Bianchi identities in the system of equations (\ref{eq:EoMBianchi0sympl})--(\ref{eq:EoMBianchi4sympl}). From the components of (\ref{eq:EoMBianchi1sympl}) with upper $A$--indices we see that $G_0^A = const := m_{\mathrm{RR}}^A$ (these parameters are associated with RR fluxes). Then (\ref{eq:EoMBianchi0sympl}) are just constraints among constants: $m^I_A m_{\mathrm{RR}}^A = 0 = e_{IA} m_{\mathrm{RR}}^A$. The upper components of (\ref{eq:EoMBianchi3sympl}) are solved by $G_2^A = dA_1^A$, defining the (electric) gauge potentials of the 4d theory. Then (\ref{eq:EoMBianchi2sympl}) are solved by $G_1^I = d\xi^I -  m_A^I A_{1}^A$ and $\tilde G_{1I}= d\tilde\xi_I - e_{IA}A_{1}^A$, where $\xi^I$ and $\tilde \xi_I$ are scalar fields. Finally, using also the quadratic constraint $e_{IA}m^I_B - m^I_A e_{IB}=0$ contained in (\ref{eq:QuadrConstrSympl}), from the lower components of (\ref{eq:EoMBianchi1sympl}) we find that $\tilde G_{0A}= e_{\mathrm{RR}A} -\xi^Ie_{IA} +\tilde\xi_Im^I_A$, where $e_{\mathrm{RR}A}$ are constant RR flux parameters. 

At this point the only equations we still have to deal with are eq.$\:$(\ref{eq:EoMBianchi4sympl}) and the lower components of $\:$(\ref{eq:EoMBianchi3sympl}). Employing the relations descending from the RR self-duality constraint, these will now be interpreted as EoM for the fields $\tilde\xi_I,\xi^I$ and $A_1^A$. Eq.$\:$(\ref{eq:EoMBianchi4sympl}) has already been treated along these lines, yielding eq.$\:$(\ref{eq:EoMBianchi4symplModifAgain}), which we take as the EoM for the scalars $\xi^I,\tilde \xi_I\,$. Concerning the EoM for $A_1^A$, we use (\ref{eq:constraint2-2}) and (\ref{eq:constr3-1}) to eliminate $\tilde G_{2A}, G_3^{\mathbb I}$ in the lower components of (\ref{eq:EoMBianchi3sympl}), and we get:
\be
d[\im\cl N_{AB} *F^B + \re\cl N_{AB}F^B ] - \tilde G_{0A}dB - e^{2\varphi}(\mathbb Q^T \widetilde{\mathbb M})_{A\mathbb{I}} *G_1^{\mathbb I} = 0\;,
\ee
where we introduced the modified field strengths $F^A$ containing the 2--form $B$:
\be\label{eq:ModifFieldStrp=0=qCase}
F^A \,:= \,G_2^A + G_0^A B\, =\, dA_1^A +  m_{\mathrm{RR}}^A B\;.
\ee

One can now check that precisely the equations of motion just obtained, together with the EoM for $B$ given in (\ref{eq:4dBEoM}), can be derived from the 4d action:\footnote{The term $\frac{1}{2} d(e^{-4\varphi}*dB)$ in (\ref{eq:4dBEoM}) is indeed derived from the piece of the 4d action associated with the reduction of the NSNS sector, see eq.$\:$(\ref{eq:decompositionNSsector}). This also fixes the overall normalization of $S^{(4)}_{\mathrm{RR}}$.\label{ftn:FtnKinTermB}}
\begin{eqnarray}\label{eq:FinalRRactionp=q=o}
\nnb S^{(4)}_{\mathrm{RR}}\!\! &=&\!\! \int_{M_4} \Big[\;\frac{1}{2}\im \cl N_{AB} F^A \wedge * F^B + \frac{1}{2}\re \cl N_{AB} F^A \wedge F^B + \frac{e^{2\varphi}}{2} \widetilde{\mathbb M}_{\mathbb{IJ}} D\xi^{\mathbb I} \wedge * D\xi^{\mathbb J}\\ [1mm]
\nnb &&\;\; +\; \frac{1}{2}dB\wedge \big[ \xi^{\mathbb I}\mathbb S_{-\mathbb I\mathbb J} D\xi^{\mathbb J} + (2 e_{\mathrm{RR}A} -\xi^I e_{IA} + \tilde \xi_I m^I_A) A_1^A\big] - \frac{1}{2} m_{\mathrm{RR}}^A e_{\mathrm{RR}A} B\wedge B \\ [1mm]
&& \;\; -\; \cl V_{\mathrm{RR}} *1 \;\Big]\;,
\end{eqnarray}
where $\xi^{\mathbb I} = {{ \xi^I} \choose {\tilde\xi_I}}$, and we have introduced the covariant derivatives
\be
\label{eq:CovDerAxionsSU3} D\xi^I \equiv G_1^I= d\xi^I -  m_A^I A_{1}^A \qquad ,\qquad D\tilde\xi_I \equiv \tilde G_{1I} = d\tilde\xi_I - e_{IA}A_{1}^A \;.
\ee
Furthermore we defined:
\be\label{eq:DefV_RR}
\cl V_{\mathrm{RR}} = -\frac{e^{4\varphi}}{2}G_0^{\mathbb A} \widetilde{\mathbb N}_{\mathbb{AB}} G_0^{\mathbb B}\;,
\ee
corresponding to the non-negative contribution of the RR sector to the scalar potential of the reduced theory.\footnote{Notice that (\ref{eq:DefV_RR}) contains a term $-\frac{e^{4\varphi}}{2}{m_{\mathrm{RR}}  \choose e_{\mathrm{RR}}}^T\widetilde{\mathbb N}{m_{\mathrm{RR}}  \choose e_{\mathrm{RR}}}$ which does not depend on the RR scalars $\xi^I$, $\tilde \xi_I$ and indeed does not contribute to their EoM. We have added it as the natural completion of the expression for $\cl V_{\mathrm{RR}}$ directly reconstructed from these EoM. The correctness of (\ref{eq:DefV_RR}) can also be verified studying the reduced Einstein equations.}

Since it yields the correct reduced EoM, we interpret the action (\ref{eq:FinalRRactionp=q=o}) as the one for the reduced type IIA RR sector. To check that $S^{(4)}_{\mathrm{RR}}$ reproduces the EoM written above, one needs the consistency constraints (\ref{eq:QuadrConstrSympl}) as well as the condition $m^I_A m_{\mathrm{RR}}^A = 0 = e_{IA} m_{\mathrm{RR}}^A$.

As mentioned above, the present setting with $p_I^A = 0= q^{IA}$ is relevant for SU(3) structure compactifications, once the specific basis of forms (of pure degree) defined e.g. in \cite{GLW1, MinasianKashani} is adopted. In this basis the parameters $e_{Ia}, m^I_a,\;a=1,\ldots b^+$, are `geometric charges', while $e_{I0}, m^I_0$ are associated with the NS flux $H^{\mathrm{fl}}$. Indeed, the action (\ref{eq:FinalRRactionp=q=o}), which has the features of an $N=2$ gauged supergravity, is in agreement with all the previous studies of $N=2$ type IIA compactifications on SU(3) structures \cite{GurLouisMicuWaldr, GaugingHeisenberg, GLW1, HousePalti, MinasianKashani, KashaniPoorNearlyKahler}. In particular, the Killing prepotentials describing the general gauging were found in \cite{GLW1} via a reduction of the gravitino susy transformations. 

It can be useful to see how several particular cases already described in the literature can be recovered. Let's take $m_{\mathrm{RR}}^A=0$ first. In this case the 2--form $B$ can be dualized to a scalar $a$. The terms in the action (\ref{eq:FinalRRactionp=q=o}) containing $dB$, together with the kinetic term $-\frac{1}{4}\int e^{-4\varphi} dB\wedge * dB$ coming from the NSNS sector (see eq.$\:$(\ref{eq:decompositionNSsector})), are then replaced by:
\be\label{eq:ActionForAxion}
S_{\mathrm{dual}}= \int_{M_4} -\frac{e^{4\varphi}}{4}\big(Da - \xi^{\mathbb I}\mathbb S_{-\mathbb I\mathbb J} D\xi^{\mathbb J} \big)\wedge * \big(Da - \xi^{\mathbb I}\mathbb S_{-\mathbb I\mathbb J} D\xi^{\mathbb J} \big)\;,
\ee
where 
\be\label{eq:covDerAxion}
Da=da -(2 e_{\mathrm{RR}A} -\xi^I e_{IA} + \tilde \xi_I m^I_A) A_1^A\;.
\ee 
The term (\ref{eq:ActionForAxion}) contributes to define a hypermultiplet quaternionic $\sigma$--model analogous to the one featured by the standard $N=2$ effective action derived from Calabi--Yau dimensional reductions \cite{FerraraSabharwal}. More specifically, the (RR sector of the) $N=2$ supergravity obtained from proper Calabi-Yau compactifications with no fluxes \cite{BodnerCadavidFerrara} is recovered by setting all the charges $e_{IA}, m^I_A, e_{\mathrm{RR}A}$ (as well as $m_{\mathrm{RR}}$) to zero. This is consistent with the fact that all the basis forms (\ref{eq:Sigma+-AsVectors}) are then closed. Allowing for non-vanishing $e_{I0},m^I_0$ yields the Calabi-Yau effective action in the presence of NS fluxes described in \cite{LouisMicu}.\footnote{With respect to \cite{LouisMicu}, we have a sign difference in the definition of the RR scalars $\tilde\xi$.}

Adopting a four dimensional approach, the $N=2$ supergravity containing $e_{\mathrm{RR}A}$, $e_{IA}$ and $m^I_A$ was obtained in \cite{GaugingHeisenberg} by performing a gauging of the Calabi-Yau effective action. The Killing vectors parameterizing the quaternionic isometries that are gauged are
\be 
k_A = (2 e_{\mathrm{RR}A} -\xi^I e_{IA} + \tilde \xi_I m^I_A)\partial_a + m_A^I\partial_{\xi^I} + e_{IA}\partial_{\tilde\xi_I} \;,
\ee
and the usual differentials $d\xi^I, d\tilde\xi_I, da$ are replaced by the covariant derivatives (\ref{eq:CovDerAxionsSU3}), (\ref{eq:covDerAxion}), coupling the scalars to the gauge vectors $A_1^A$.

Furthermore, taking just $e_{0A}\neq 0$, we find agreement with the results of \cite{GurLouisMicuWaldr} for type IIA reductions on half-flat manifolds (the parameter $e_{00}$ being associated with an NS flux). 

Finally, let's consider nonvanishing $m^A_{\mathrm{RR}}$. These parameters generate some couplings for the NS 2--form $B$, including a mass term: then $B$ cannot be dualized to an axion \cite{LouisMicu, GLW1}. If $m^I_A =0 = e_{IA} $,  eq.$\:$(\ref{eq:FinalRRactionp=q=o}) precisely reproduces the RR part of the action derived in \cite{LouisMicu} for Calabi-Yau compactifications of type IIA with RR fluxes.

\subsection{General case}

Let's consider a general charge matrix $\mathbb Q$ as given in (\ref{eq:ChargeMatrixQ}). An $N=2$ lagrangian including this same set of charges was obtained in \cite{D'AuriaFerrTrigiante} using purely 4d supergravity techniques and building on results in \cite{GaugingHeisenberg, N=2withTensor1, N=2withTensor2}. Having the $N=2$ effective theory arising from Calabi-Yau compactifications as a starting point, the authors of \cite{D'AuriaFerrTrigiante} first deformed it by implementing a standard electric gauging of the quaternionic isometries, and subsequently performed a dualization of a subset of the RR axions to antisymmetric 2--tensors in order to include the magnetic charges.

In section$\:$\ref{ReductionNSNS} we found consistency between this procedure and the dimensional reduction of the NSNS sector, obtaining in particular eq.$\:$(\ref{eq:VNSinTermsOf4dFields}) for the NSNS scalar potential. Here we approach the same question for the RR sector. As in the previous subsection, we construct a 4d action via the analysis of the reduced RR EoM/Bianchi identities. A set of 2--form potentials, beside the vector and scalar fields, will emerge directly from the analysis of the selected 4d Bianchi identities. The outcome of the analysis is summarized in Table$\:$\ref{OutcomeRReqs}.

Even if for a general $\mathbb Q$ all the equations (\ref{eq:EoMBianchi0sympl})--(\ref{eq:EoMBianchi4sympl}) are symplectically covariant, we will anyway break this symmetry in order to establish a set of EoM associated with a 4d action written in terms of electric vectors only. For this task we introduce appropriate projectors that we will apply to eqs.$\:$(\ref{eq:EoMBianchi0sympl})--(\ref{eq:EoMBianchi4sympl}). In the following computations, several technical steps are close to the ones employed in \cite{D'AuriaFerrTrigiante} for the dualization of the RR axions to antisymmetric 2--tensors.

We start splitting the charge matrix $\mathbb Q$ in the following $(2b^-+2)\times (b^+ + 1)$ submatrices:
\be\label{eq:DefSubmatrOfQ}
U^{\mathbb I}_{\;A} := \mathbb Q^{\mathbb I}_{\;A} = {m^I_{\;A}  \choose {e_{IA}}}\qquad,\qquad V^{\mathbb IA} := \mathbb Q^{\mathbb IA} = {q^{IA}  \choose {p_{I}^{\;A}}}\;.
\ee
With respect to the gauge vectors with upper indices $A_1^A$ that we are going to define below, the elements of $U$ are electric charges, while $V$ contains magnetic charges. 

As in \cite{D'AuriaFerrTrigiante}, we adopt the working assumptions $b^+ \leq b^-$, and that the matrix $U$ has maximal rank $b^++1$. Then we introduce the matrix $\widetilde U^{\!A}_{\;\;\,\mathbb I}\,$, defined through:
\be\label{eq:DefUtilde}
\widetilde U^{\!A}_{\;\;\mathbb I}U^{\mathbb I}_{\;B}= \delta^A_{\;B}\qquad,\qquad U^{\mathbb I}_{\;A} \widetilde U^{\!A}_{\;\;\mathbb J}= (\mathbb P_{\neq 0})^{\!\mathbb I}_{\;\mathbb J}\;,
\ee
$\mathbb P_{\neq 0}$ being the projector on the subspace corresponding to the non-vanishing minor of $U^{\mathbb I}_{\;A}$. We also define the orthogonal projector $(\mathbb P_{ 0})^{\!\mathbb I}_{\;\mathbb J} \equiv \delta^{\mathbb I}_{\;\mathbb J}\, -\, (\mathbb P_{\neq 0})^{\!\mathbb I}_{\;\mathbb J}\,$.

An identity we will need is:
\be\label{eq:identityForqp}
V  \;=\; V  U^T\widetilde U^T \;=\; U V^T\widetilde U^T \;,
\ee
which is obtained recalling the first of (\ref{eq:DefUtilde}) and then the first of (\ref{eq:QuadrConstrSympl}). Notice that $(\widetilde UV)^{AB}$ is then symmetric.

\vskip .4cm

\underline{\emph{Bianchi identities and fundamental 4d fields}}\\ [-2mm]

\begin{table}
\begin{center}
\begin{tabular}{|c|c||c|c|}
\hline
Equation                    & 	yields:   & Equation                      & yields: \\
\hline\hline
\rule{0pt}{2.5ex} (\ref{eq:EoMBianchi0sympl}) & constraints among charges &(\ref{eq:EoMBianchi3sympl}) &Bianchi for $G_2^A$ $\rightarrow \;$def.$\,A^A_1$ \\
\rule{0pt}{0ex}   (\ref{eq:EoMBianchi1sympl}) &     expression for $G_0^{\mathbb A}$      &                               &  EoM for $A_1^A$  \\
\hline
\rule{0pt}{2.7ex} (\ref{eq:EoMBianchi2sympl}) &   Bianchi for $\widehat G_1^{\mathbb I}$ $\rightarrow$ def.$\;\,\widehat \xi^{\:\mathbb I}$& (\ref{eq:EoMBianchi4sympl}) & Bianchi for $\check G_{3A}$ $\rightarrow \;$def.$\,\check C_{2A}$ \\ 
                  & EoM for $\check C_{2A}$ & (rewr. as (\ref{eq:EoMBianchi4symplModifAgain})) & EoM for $\widehat \xi^{\:\mathbb I}$\\
\hline
\end{tabular}
\caption{Analysis of the reduced RR equations for a general charge matrix $\mathbb Q$.} \label{OutcomeRReqs}
\end{center}
\end{table}

\noindent With respect to the analysis of subsection$\:$\ref{p=0=qcase}, the presence of the $p^A_I$ and $q^{IA}$ charges makes less trivial the identification and the solution of a set of Bianchi identities for the fundamental 4d fields. For this purpose we make use of the matrices defined here above. As we will see, a set 2--form degrees of freedom will be required.

We start introducing a set of scalar fields. Define \cite{D'AuriaFerrTrigiante}:
\be\label{eq:DefCheckAndHatG1}
\check G_1^A := \widetilde U^{\!A}_{\;\;\mathbb I}G_1^{\mathbb I} \qquad,\qquad  \widehat G_1^{\mathbb I} := {\mathbb P_{0}}^{\!\mathbb I}_{\;\mathbb J}G_1^{\mathbb J}\;, \ee
so that
\be\label{eq:decompositionG1}
G_1^{\mathbb I} = U^{\mathbb I}_{\;A}\check G_1^A + \widehat G_1^{\mathbb I}\;.
\ee
We want to keep the $\widehat G_1^{\mathbb I}$, while we will deal with $\check G_1^A$ in the next paragraph. We act with $\mathbb P_{0}$ on eq.$\:$(\ref{eq:EoMBianchi2sympl}) and we observe that $\mathbb P_{0}\mathbb Q = 0\,$, due to the definition of $\mathbb P_{0}$ below eq.$\:$(\ref{eq:DefUtilde}) and to identity (\ref{eq:identityForqp}). Then we get
\be\label{eq:SolHatG1}
d \widehat G_1^{\mathbb I} = 0\qquad \Rightarrow \qquad \widehat G_1^{\mathbb I} = d \widehat \xi^{\:\mathbb I}\;,
\ee
with $\widehat \xi^{\:\mathbb I}$ being a set of real scalars satisfying $(\mathbb P_{\neq 0})^{\mathbb I}_{\;\:\mathbb J}\,\widehat \xi^{\:\mathbb J} =0$ and corresponding therefore to $\mathrm{rank}(\mathbb P_{0})= 2(b^-+1)- (b^+ + 1)$ degrees of freedom.

Recalling (\ref{eq:decompositionG1}) and (\ref{eq:QuadrConstrSympl}), eq.$\:$(\ref{eq:EoMBianchi1sympl}) can then be written as 
\be\label{eq:ResultForG0}
d G_0^{\mathbb A} - \widetilde{\mathbb Q}^{\mathbb A}_{\;\;\,\mathbb I} d\widehat \xi^{\:\mathbb I}  = 0\qquad\Rightarrow \qquad 
G_0^{\mathbb A} \; =\;  c^{\mathbb A}  +  \widetilde{\mathbb Q}^{\mathbb A}_{\;\;\,\mathbb I} \widehat \xi^{\:\mathbb I} \;,
\ee
with $c^{\mathbb A} = { {m_{\mathrm{RR}}^A} \choose {e_{\mathrm{RR}A}} }$ a vector of constant charges, associated with general RR background fluxes. Again employing (\ref{eq:QuadrConstrSympl}), eq.$\:$(\ref{eq:EoMBianchi0sympl}) translates in the following consistency condition among the different parameters \cite{D'AuriaFerrTrigiante}:
\be\label{eq:constrRRandGeomCharges}
\mathbb Q^{\mathbb I}_{\;\,\mathbb A} c^{\mathbb A} = 0\;.
\ee

Next we define the $b^++1$ combinations
\be\label{eq:DefCheckG3}
\check G_{3A}:=\, -(U^T\mathbb S_-)_{A\mathbb I} G_3^{\mathbb I} \,.
\ee
Multiplying eq.$\:$(\ref{eq:EoMBianchi4sympl}) by $U^T\mathbb S_-$ from the left, and recalling (\ref{eq:QuadrConstrSympl}), we get 
\be\label{eq:BianchiCheckG3}
d\check G_{3A}=0\;,
\ee
which we choose to solve as
\be\label{eq:ExprCheckG3}
\check G_{3A}= d(\check C_{2A}+ \zeta_A B) \,,
\ee
where the 2--forms $\check C_{2A}$ are new fields, $B$ is the NS 2--form and $\zeta_A$ is a combination of the scalars $\widehat \xi^{\:\mathbb I}$ to be specified below. The 2--forms $\check C_{2A}$ will be dynamical fields of our eventual 4d action.

Let's finally turn to gauge vectors. Here we choose to define fundamental vector potentials with upper indices only, so we keep all the $G_2^A$ and dualize all the $\tilde G_{2A}$, breaking in this way the symplectic structure for the 2--forms $G_2^{\mathbb A}$. The components of (\ref{eq:EoMBianchi3sympl}) with upper indices can be read as Bianchi identities for $G_2^A$, while the dualization of the lower components will provide the EoM for the associated vector potentials. First we look at the Bianchi identities, which read:
\be\label{eq:BianchiG2}
dG_2^A + (V^T\mathbb S_-)^{\!A}_{\;\;\mathbb I}G_3^{\mathbb I} =0\;.
\ee
Using (\ref{eq:identityForqp}) and (\ref{eq:DefCheckG3}), we rewrite this as $ dG_2^A - (\widetilde UV)^{A B}  \check G_{3B} =0$. Taking (\ref{eq:ExprCheckG3}) into account, this last equation is solved introducing a set of vector potentials $A_{1}^A$:
\be\label{eq:SolForG_2}
G_2^A\; =\; d A_{1}^A + (\widetilde UV)^{A B} (\check C_{2B} + \zeta_B B)\;.
\ee
We now fix the $\zeta_A$ introduced in (\ref{eq:ExprCheckG3}). We choose
\be
\zeta_A \equiv (U^T\mathbb S_-)_{A\mathbb I}\widehat\xi^{\:\mathbb I}\,,
\ee 
in such a way that the $b^+ + 1$ two--forms
\be\label{eq:Def4dModifFieldStr}
F^A \::=\: G_2^A + G_0^AB \:=\: dA_1^A+  (\widetilde UV)^{A B} \check C_{2B} +  m_{\mathrm{RR}}^A B
\ee
contain vectors and 2--form potentials only (to obtain this expression recall also (\ref{eq:identityForqp}) and (\ref{eq:ResultForG0})). Thus the $F^A$ are a set of field strengths for the vector potentials $A_1^A$, modified by the presence of the 2--forms $B, \check C_2^A$, and generalize the field strengths (\ref{eq:ModifFieldStrp=0=qCase}) to the case of nonvanishing $V^{\mathbb IA}$ charges. These are the appropriate modified field strengths described by the formalism of $N=2$ supergravity with tensor multiplets\footnote{Notice that one could also express the $\check C_{2A}$ by introducing a redundant set of $2b^-+2$ two-forms $C_2^{\mathbb I}= {{C_2^I} \choose {\tilde C_{2I}}}$ and writing, in analogy with (\ref{eq:DefCheckG3}), $\check C_{2A}= -(U^T\mathbb S_-)_{A\mathbb I} C_2^{\mathbb I}= C_2^I e_{IA} - \tilde C_{2I} m^I_A$. Then, recalling (\ref{eq:identityForqp}), eq.$\:$(\ref{eq:Def4dModifFieldStr}) would become $F_2^A = d A_{1}^A + C_2^I p^A_I  - \tilde C_{2I}q^{IA} + m_{\mathrm{RR}}^AB$. However the only propagating degrees of freedom would be just the combinations of $C_2^I$ and $\tilde C_{2I}$ associated with $\check C_{2A}$ \cite{GLW2, D'AuriaFerrTrigiante}. Analogously, as in subsection$\:$\ref{p=0=qcase} we could introduce a symplectic vector of $2b^-+2$ scalars ${\xi}^{\mathbb I}={\xi^I  \choose \tilde \xi_I}$ such that $\widehat \xi^{\:\mathbb I} = {\mathbb P_{0}}^{\!\mathbb I}_{\;\mathbb J}{\xi}^{\mathbb J}\,$. Then the result of (\ref{eq:ResultForG0}) would read: $G_0^A = m_{\mathrm{RR}}^A+\xi^Ip_I^A-\tilde\xi_Iq^{IA}$ and $\tilde G_{0A}= e_{\mathrm{RR}A} -\xi^Ie_{IA} +\tilde\xi_Im^I_A$. However, in these expressions the only relevant combinations of the $\xi^{\mathbb I}$ correspond to the $\widehat \xi^{\:\mathbb I}$.} \cite{N=2withTensor1, N=2withTensor2, DeWitSamtlebenTrig, D'AuriaFerrTrigiante}.

To summarize, the outcome of this paragraph is a set of fundamental degrees of freedom $\widehat\xi^{\:\mathbb I}$, $\check C_{2A}$ and $A_1^A$, related to $\widehat G_1^{\mathbb I}$, $\check G_{3A}$ and $G_2^A$ as in (\ref{eq:SolHatG1}), (\ref{eq:ExprCheckG3}) and (\ref{eq:SolForG_2}). Furthermore in (\ref{eq:Def4dModifFieldStr}) we defined the proper modified field strengths for $A_1^A$, and in (\ref{eq:ResultForG0}) we expressed $G_0^{\mathbb A}$ as a combination of scalars and charges. The charges have to satisfy conditions (\ref{eq:constrRRandGeomCharges}).

\vskip .4cm

\underline{\emph{Equations of motion}}\\ [-2mm]

\noindent We now establish the EoM associated with the identified fundamental 4d fields. For this purpose we study the projections of eqs.$\:$(\ref{eq:EoMBianchi0sympl})--(\ref{eq:EoMBianchi4sympl}) which are independent with respect to the ones considered in the above study of the Bianchi identities.

The EoM for the vector potentials $A_1^A$ are obtained from the lower components of (\ref{eq:EoMBianchi3sympl}) using the duality relation (\ref{eq:constraint2-2}) to eliminate $\tilde G_{2A}$, recalling expressions (\ref{eq:DefCheckG3}), (\ref{eq:ExprCheckG3}) as well as the definition of $F^A$ in (\ref{eq:Def4dModifFieldStr}), and noticing that $\tilde G_{0A}= e_{\mathrm{RR}A}+ \zeta_A$. The result is:
\be\label{eq:EoMAgeneralCase}
d\big(\im \mathcal N_{AB} *F^B + \re\mathcal N_{AB} F^B + \check C_{2A} - e_{\mathrm{RR}A} B \big) =0\;.
\ee
\vskip .2cm
Next we find an expression for the $\check G_1^A\,$ defined in (\ref{eq:DefCheckAndHatG1}). Multiplying relation (\ref{eq:constr3-1}) by $U^T\mathbb S_-$ from the left, substituting (\ref{eq:decompositionG1}) in it and recalling (\ref{eq:QuadrConstrSympl}), (\ref{eq:DefCheckG3}) as well as the expressions for $\check G_{3A}$, $\widehat G_1^{\mathbb I}$ and $\zeta_A$ obtained in the study of the Bianchi identities, we arrive at:
\be\label{eq:expressG1A}
\check G_1^A = - \Delta^{-1\,AB} \big[*d\check C_{2B}+ \zeta_B *dB + e^{2\varphi}( U^T\widetilde{\mathbb M})_{B\mathbb I}\,d\widehat\xi^{\:\mathbb I}\,\big] \,,
\ee
where we introduced the symmetric matrix \cite{D'AuriaFerrTrigiante}:
\be
\Delta_{AB}:= e^{2\varphi}(U^T)_A^{\;\;\,\mathbb I}\, \widetilde{\mathbb M}_{\mathbb I\mathbb J} \, U^{\mathbb J}_{\;\; B}\;.
\ee

In order to get the EoM associated with $\check C_{2A}$, we start acting with $\widetilde U$ from the left on eq.$\:$(\ref{eq:EoMBianchi2sympl}) and exploiting (\ref{eq:constraint2-2}) in order to eliminate $\tilde G_{2A}$. After some steps involving the expressions arising from the Bianchi identites, we obtain
\be\label{eq:EoMC2GeneralCase}
d \check G_1^A  +  dA_1^A  +  (\widetilde U V)^{AB} \big[\im \mathcal N_{BC} *F^C + \re\mathcal N_{BC} F^C + \check C_{2B} -  e_{\mathrm{RR}B} B \big]\, =\, 0\,,
\ee
where $\check G_1^A$ should be read as (\ref{eq:expressG1A}).
\vskip .2cm
The EoM for the scalars $\widehat \xi^{\:\mathbb I}$ are obtained substituting (\ref{eq:decompositionG1}) in (\ref{eq:EoMBianchi4symplModifAgain}) and lowering the symplectic index with $\mathbb S_-$:
\bea
\nnb -\, d\big[\,e^{2\varphi} \widetilde{\mathbb M}_{\mathbb I\mathbb J} *(d\widehat \xi^{\:\mathbb J} + U^{\mathbb J}_{\;\,A}\check G_1^A)\,\big] \,+\, dB\wedge [\,(\mathbb S_- U)_{\mathbb IA}\check G_1^A + (\mathbb S_-d\widehat \xi\,)_{\mathbb I}\,]\qquad && \\ [1mm]
\label{eq:EoMHatXi} -\, e^{4\varphi}(\mathbb S_-\mathbb Q \mathbb {N})_{\mathbb{IA}}G_0^{\mathbb A} *1 &=& 0\,,
\eea
where again expression (\ref{eq:expressG1A}) for $\check G_1^A$ should be substituted. Once this is done\footnote{Taking into account the explicit expression for $\check G_1^A$, one can see that the $b^++1$ linear combinations of the equations (\ref{eq:EoMHatXi}) obtained via multiplication by $(U^T)_{\!A}^{\;\,\mathbb I}$ vanish identically, as it should be: we have already exploited these combinations to write (\ref{eq:BianchiCheckG3}).}, the piece of (\ref{eq:EoMHatXi}) associated with a kinetic term for the $\widehat \xi^{\:\mathbb I}$ reads $- d\big( \widetilde\Delta_{\mathbb I\mathbb J} *d\widehat \xi^{\:\mathbb J} \big)$, with \cite{D'AuriaFerrTrigiante}:
\be
\widetilde\Delta_{\mathbb I\mathbb J} = e^{2\varphi}\big(\widetilde{\mathbb M} - e^{2\varphi}\widetilde{\mathbb M}U\Delta^{-1}U^T \widetilde{\mathbb M}\big)_{\mathbb I\mathbb J}  \;.
\ee
\vskip .2cm
Finally, we rewrite the EoM for the four dimensional $B$--field given in (\ref{eq:4dBEoM}) substituting the expressions for the fundamental 4d fields. After some steps we arrive at:
\bea
\nnb \frac{1}{2} d(e^{-4\varphi}*dB)\, + \, m_{\mathrm{RR}}^A\big(\im \mathcal N_{AB} *F^B + \re\mathcal N_{AB} F^B \big) \,-\, e_{\mathrm{RR}A} F^A \qquad&&\\
\label{eq:EoMBGeneralCase} -\, \frac{1}{2}d\widehat\xi^{\:\mathbb I}\mathbb S_{-\mathbb I\mathbb J}d\widehat\xi^{\:\mathbb J}\, +\, d(\zeta_A\check G_1^A) &=&0\;.
\eea

\vskip .4cm

\underline{\emph{4d action for the reduced RR sector}}\\ [-2mm]

\noindent We can now reconstruct the action yielding the EoM (\ref{eq:EoMAgeneralCase}), (\ref{eq:EoMC2GeneralCase}), (\ref{eq:EoMHatXi}) and (\ref{eq:EoMBGeneralCase}), respectively associated with the fields $A_1^A$, $\check C_{2A}$, $\widehat\xi^{\:\mathbb I}$ and $B$ (for this last remind footnote \ref{ftn:FtnKinTermB}). We find:
\bea
\nnb S_{\mathrm{RR}}^{(4)}\!\!\! &=&\!\!\! \int_{M_4} \Big\{\;\frac{1}{2}\im \cl N_{AB} F^A \wedge * F^B + \frac{1}{2}\re \cl N_{AB} F^A \wedge F^B + \frac{1}{2} \widetilde{\Delta}_{\mathbb{IJ}} d\widehat\xi^{\:\mathbb I} \wedge * d\widehat \xi^{\:\mathbb J}\\ [2mm]
\nnb &&\;\; +\; \frac{1}{2}\Delta^{-1AB} (d\check C_{2A} + \zeta_A  dB)\wedge * (d\check C_{2B} + \zeta_B  dB)  \\ [2mm]
\nnb &&\;\; +\; (d\check C_{2A} +  \zeta_A dB)\wedge(e^{2\varphi}\Delta^{-1}U^T \widetilde{\mathbb M})^{\!A}_{\;\,\mathbb I}d\widehat \xi^{\:\mathbb I} \,+\, \frac{1}{2}dB \wedge \widehat\xi^{\:\mathbb I}\,\mathbb S_{-\mathbb I\mathbb J}d \widehat\xi^{\:\mathbb J} \\ [2mm] 
\label{eq:RRactionGeneralCase}&&\;\; +\; (\check C_{2A}- e_{\mathrm{RR} A} B)\wedge \big [dA_1^A + \frac{1}{2}(\widetilde UV)^{AB} \check C_{2B} + \frac{1}{2}m_{\mathrm{RR}}^A B \big] \; -\; \cl V_{\mathrm{RR}} *1 \Big\}\,.
\eea
In order to derive the EoM, constraint (\ref{eq:constrRRandGeomCharges}) (written in the form $Um_{\mathrm{RR}} + Ve_{\mathrm{RR}}= 0$) should be recalled. The RR contribution to the 4d scalar potential is defined as in (\ref{eq:DefV_RR}):
\be\label{eq:DefV_RRbis}
\cl V_{\mathrm{RR}} = -\frac{e^{4\varphi}}{2}G_0^{\mathbb A} \widetilde{\mathbb N}_{\mathbb{AB}} G_0^{\mathbb B}\;,
\ee
but in the present general case expression (\ref{eq:ResultForG0}) for $G_0^{\mathbb A}$ should be used. Using (\ref{eq:mathbbMandmathbbN}), eq.$\:$(\ref{eq:DefV_RRbis}) can be derived from the geometric formula
\be\label{eq:GeomFormulaForVRR}
\cl V_{\mathrm{RR}} \,=\, \frac{e^{4\varphi}}{2}\int_{M_6}\langle G , *_b G\rangle\;,
\ee 
where $G:= G_0^A\om_A - \tilde G_{0A}\tilde\om^A$ corresponds to the purely internal part of the RR field $\bf{\hat G}$, expanded as in (\ref{eq:ExpansionWithG}). This is a non-negative expression. 

Notice that $\cl V_{\mathrm{RR}}$ effectively vanishes when integrating out the subset of the scalars $\widehat \xi^{\:\mathbb I}$ entering in the potential \cite{D'AuriaFerrTrigiante}: indeed from the $\widehat \xi^{\:\mathbb I}$--EoM (\ref{eq:EoMHatXi}) evaluated in a vacuum one gets the condition $G_0^{\mathbb A}=0$.

The dimensionally reduced action (\ref{eq:RRactionGeneralCase}) coincides with the one found in \cite{D'AuriaFerrTrigiante} using purely four dimensional $N=2$ supergravity techniques. It contains topological as well as mass terms for the 2--forms $B$ and $\check C_{2A}$, with mass matrix:
\be
M^2 = -\left( \begin{array}{cc}  m_{\mathrm{RR}}^T\im\cl N m_{\mathrm{RR}}  &  m_{\mathrm{RR}}^T\im\cl N \widetilde{U}V  \\
(\widetilde{U}V)^T\im\cl N m_{\mathrm{RR}}   &  (\widetilde{U}V)^T\im\cl N \widetilde{U}V \end{array}\right)\;.
\ee

\section{Discussion}\label{discussion}

\setcounter{equation}{0}

Joining the results for the reduced NSNS and RR sectors, derived in sections \ref{ReductionNSNS} and \ref{ReductionRRsector} respectively, we get the complete bosonic action associated with $N=2$ flux compactifications of type IIA supergravity on \sst structures. 

This $N=2$ supergravity involves massive tensor multiplets, and is in agreement with the one that ref.$\:$\cite{D'AuriaFerrTrigiante} obtained starting from the Calabi-Yau 4d effective action, gauging the Heisenberg algebra of quaternionic isometries and then dualizing a set of axions in order to introduce the magnetic charges.
In our approach to the reduction of the RR sector we didn't need to perform any {\it a posteriori} dualization of scalars: reducing the RR EoM/Bianchi identities we identified and solved a set of 4d Bianchi identities already encoding the appropriate degrees of freedom.

The application of the generalized geometry formalism allowed to derive a geometric formula for the full 4d scalar potential $\cl V= \cl V_{\mathrm{NS}} + \cl V_{\mathrm{RR}}$, given by eqs.$\:$(\ref{eq:V_NSpureSpinors}) and (\ref{eq:GeomFormulaForVRR}). Expanding the pure spinors as well as the internal RR field strengths on the basis polyforms, and integrating over the compact manifold, we recover the symplectically invariant scalar potential of \cite{D'AuriaFerrTrigiante}. The NSNS contribution to the potential is mirror symmetric under the exchange $\Phi_+\leftrightarrow \Phi_-$, while we expect the type IIB RR contribution still read as (\ref{eq:GeomFormulaForVRR}), modulo the substitution $G^{\mathrm{even}}\to G^{\mathrm{odd}}$.

Our expression for the potential is also relevant when considering \hbox{$N=2\to N=1$} truncations, for instance induced by orientifold planes. Indeed one can get the $N=1$ scalar potential via a reformulation of the $N=2$ potential in terms of the appropriate $N=1$ variables (in the context of generalized geometry these were first derived in \cite{GrimmBen}), in the same way as the $N=1$ superpotential and D-terms can be obtained from the Killing prepotentials defining the $N=2$ gaugings \cite{GLW1, GLW2, CassaniBilal}. It should also be possible to derive the expression for the $N=1$ scalar potential including the effects of a non-trivial warp factor, along the lines of \cite{KoerberMartucci10to4}. Indeed our expression (\ref{eq:FormulaForInternalNSsector}), reformulating the internal NSNS sector in terms of the generalized geometry data, can in principle be extended to take the warping into account.

Concerning the basis forms defining the truncation, it would be interesting to start from some well-characterized class of internal manifolds with \sst structure and exhibit an explicit construction. In particular, it would be nice to find an example in which the basis defining the truncation is provided by forms of mixed degree.

A better characterization of the expansion forms could also allow to conclude about the consistency of the truncation, for instance checking whether the 4d solutions lift to 10d solutions (see \cite{KashaniPoorNearlyKahler} for a first example in this sense). In this context, in order to study the 10d Einstein equations it would be useful to dispose of a formula generalizing (\ref{eq:FormulaForInternalNSsector}) and expressing the full Ricci tensor of the internal manifold, and not just its trace, in terms of the \sst structure data.

\vskip 1cm

{\large \bf Acknowledgments}\\ [2mm]
I am grateful to Adel Bilal for many illuminating discussions, advice, and comments on the manuscript. I would also like to thank Mariana Gra{\~n}a, Luca Martucci and \hbox{Alessandro} Tomasiello for conversations and/or correspondence. This work is supported in part by the EU grants MRTN-CT-2004-005104 and MRTN-CT-2004-512194, by the French ANR grant ANR(CNRS-USAR) no.05-BLAN-0079-01 as well as by the ``Programme Vinci 2006 de l'Universit\'e Franco-Italienne''.


\appendix

\section{Conventions}\label{conventions}

\setcounter{equation}{0}

\subsection{Hodge dual}

In the main text we deal with a $M_{10}= M_4\times M_6$ spacetime. $M_6$ is a Riemannian manifold, while $M_{10}$ and $M_4$ are Lorentzian manifolds with a mostly + signature metric: $(-+\ldots +)$. 

Our definition of the Hodge dual on $M_d$ is:
\be\label{eq:HodgeStar}
*(dx^{\mu_1}\wedge\ldots\wedge dx^{\mu_p}):= \frac{1}{(d-p)!}\epsilon^{\mu_1\ldots \mu_p}_{\phantom{\mu_1\ldots \mu_p}\mu_{p+1}\ldots \mu_d} dx^{\mu_{p+1}}\wedge\ldots\wedge dx^{\mu_d} \;,
\ee
with $\epsilon_{12\ldots d}= \sqrt{|g_d|}$. In the main text the $x^{\mu}$ coordinates are associated with $M_4$, but in (\ref{eq:HodgeStar}) and in the forthcoming (\ref{eq:contraction}) they are generic for $M_d$. We recall that on a $p$--form $A_p\,$:
\be
** A_p = (-)^{p(d-p)+t}A_p\;,
\ee
where $t=0$ if $M_d$ is Riemannian, and $t=1$ if $M_d$ is Lorentzian.

If $A_p$ and $B_q$ are $p$-- and $q$-- forms respectively ($p\leq q$) we define
\be\label{eq:contraction}
A_p\lrcorner B_q := \frac{1}{p!(q-p)!} A^{\mu_1\ldots \mu_p}B_{\mu_1\ldots \mu_p \mu_{p+1}\ldots \mu_q}dx^{\mu_{p+1}}\wedge\cdots\wedge dx^{\mu_q} \;.
\ee
Then we have
\be\label{eq:*becomesContract}
A_p\wedge * B_p = A_p\lrcorner B_p *1 \;,
\ee
so that the kinetic term of a $p$--form potential $A_p$ can be written as $ -\frac{1}{2}\int dA\wedge *dA\,$.

If $\hat F_p = F_{p-k}\wedge \om_k $ is a $p$--form living on $M_{10}$, while $F_{p-k}$ lives on $M_4$ and $\om_k$ lives on $M_{6}$, then the 10d Hodge dual splits into 4d and 6d Hodge duals as follows:
\be
* \hat F_n = (-1)^{k(n-k)}* F_{n-k}\wedge * \om_k\;.
\ee
Recalling the definition of the involution $\lambda$ in eq.$\:$(\ref{eq:10dSelfDualityF}) we also deduce
\be\label{eq:*lambdaSplits}
*\lambda(\hat F_n) = *\lambda(F_{n-k})\wedge *\lambda(\om_k)\;.
\ee

\subsection{Gamma matrices, Spin(6) spinors and SU(3) structures}\label{SU3strConv}

The Cliff(6) gamma matrices $\gamma^m$ are all purely imaginary and hermitian. The six-dimensional chirality matrix is defined as:
\be\label{eq:chiralitygamma}
\gamma=\frac{i}{6!}\epsilon_{mnpqrs}\gamma^{mnpqrs}\;,
\ee
and the following identity holds:
\be\label{eq:chiralgammaOngammas}
\gamma\gamma_{m_1\ldots m_k}= \frac{i(-)^{[\frac{k+1}{2}]}}{(6-k)!}\epsilon_{m_1\ldots m_k m_{k+1}\ldots m_6}\gamma^{m_{k+1}\ldots m_6}\;.
\ee
If $\eta_+$ is a Spin(6) spinor satisfying $\gamma\eta_+=\eta_+$, then we define its chiral conjugate as $\eta_-\equiv \eta_+^*$.

The bispinors introduced in the main text are better seen using the following Fierz identity between two Spin(6) spinors $\psi, \chi\,$:
\be
\label{eq:fierz} \psi\otimes \chi^\dag\;=\; \frac{1}{8}\sum_{k=0}^6\frac{1}{k!}\big( \chi^{\dag}\gamma_{m_k\ldots m_1}\psi \big)\gamma^{m_1\ldots m_k}\;.
\ee

Let's now turn to the SU(3) structure conventions. We relate the different SU(3)--invariant objects on $M_6$ as follows:
\be\label{eq:relationJgI}
g_{mn}= J_{mp}I^p_{\;\;n}\;,\ee 
\be\label{eq:DefJOmegaFromEta}
J_{mn}= \mp i\eta^\dag_\pm \gamma_{mn}\eta_\pm ||\eta_+||^{-2}\qquad,\qquad \Om_{mnp}= -i\eta_-^\dag\gamma_{mnp}\eta_+||\eta_+||^{-2} \;.
\ee
where $\eta_\pm$ are globally defined and nowhere vanishing chiral spinors, $I$ is the almost complex structure ($I^2=-id$), $J$ is the almost symplectic 2--form, and $\Om$ is the (3,0)--form. $J$ and $\Om$ satisfy $J\wedge \Om =0$, so that $J$ is (1,1) with respect to $I$. 

A useful decomposition of the chirality projectors on the basis of eigenstates $\{ \eta_{\pm}, \gamma^{m}\eta_{\mp}  \}$ is:
\be\label{eq:ChiralityAsProjector}
\frac{1 \pm \gamma}{2} = \big(\eta_\pm \eta_\pm^\dag +\frac{1}{2}\gamma^m \eta_\mp \eta_\mp^\dag \gamma_m\big) ||\eta_+||^{-2}\;.
\ee
Then one has:
\bea
\label{eq:gammaEta}\gamma_m\eta_+ &=& -iJ_{mn}\gamma^n \eta_+\\
\gamma_{mn}\eta_+ &=& i J_{mn}\eta_+ +\frac{i}{2}\Om_{mnp}\gamma^p \eta_-\\
\gamma_{mnp}\eta_+ &=& i\Om_{mnp}\eta_- + 3i J_{[mn}\gamma_{p]}\eta_+\;.
\eea
Using the holomorphic projector $P= \frac{1}{2}(1 - iI)$ we can introduce the gamma matrices with holomorphic/antiholomorphic indices $i,\bar\imath =1,2,3\,$: 
\be
\gamma^i := P^i_{\;\;n} \gamma^n\quad\textrm{and}\quad \gamma^{\bar\imath} := \bar P^{\bar\imath}_{\;\;n}\gamma^n\;.
\ee 
From (\ref{eq:gammaEta}) and (\ref{eq:relationJgI}) we see that $\gamma^{i}\eta_+ = 0$. Instead $\gamma^{\bar \imath}\eta_+$ transforms in the $\bf{\bar 3}$ of SU(3).

With the conventions listed above, one also has:
\be\label{eq:*Jand*1}
*J= \frac{1}{2} J\wedge J\qquad,\qquad *1\equiv vol_6 = \frac{1}{6}J\wedge J\wedge J= \frac{i}{8}\Om\wedge \bar \Om\;,
\ee
as well as, using (\ref{eq:fierz}):
\be\label{eq:SU3PureSpinors}
8\eta_+\otimes\eta_+^\dag = ||\eta_+||^{2}e^{-iJ}\qquad,\qquad 8\eta_+\otimes\eta_-^\dag =-i||\eta_+||^{2}\,\Om\;.
\ee

\subsection{SU(3)$\times$SU(3) structures and Spin(6) spinors}\label{sstAndSpin6}

On the bispinors $\slashPhipm = 8\eta^1_+\otimes \eta^{2\dag}_\pm$ (the Clifford map ``$\,/\,$'' was defined in (\ref{eq:DefCliffordMap})) one naturally defines an action of $\gamma^{i_1}$, $\gamma^{\bar\imath_1}$ from the left and of $\gamma^{i_2}$, $\gamma^{\bar\imath_2}$ from the right, where $\gamma^{i_1}$ ($\gamma^{i_2}$) is holomorphic with respect to the almost complex structure $I_1$ ($I_2$) associated with $\eta^1_+$ ($\eta^2_+$). Then the 6 annihilators of the pure spinor $\slashPhiPlus$ are $\buildrel\to\over\gamma\!{}^{i_1}$ and $\buildrel\leftarrow\over\gamma\!{}^{\bar \imath_2}$, while $\slashPhiMinus$ is annihilated by $\buildrel\to\over\gamma\!{}^{i_1}$ and $\buildrel\leftarrow\over\gamma\!{}^{i_2}$. The conjugate gamma matrices act as creators. Applying the Clifford map backwards, these facts can also be translated in the polyform picture. For the gamma matrices, the dictionary is \cite{GMPT3}:
\be\label{eq:gammaSlashA}
\gamma^m \slash \!\!\!\!A_\pm = \begin{picture}(10,10)(-15,5)
\put(0,0){\line(6,1){85}}\end{picture}
(dx^m\!\!\wedge + g^{mn} \iota_{\partial_n})A_\pm \qquad,\qquad \slash \!\!\!\!A_\pm \gamma^m = \pm \!\!\!\begin{picture}(10,10)(-15,5)
\put(0,0){\line(6,1){85}}\end{picture}(dx^m\!\!\wedge - g^{mn}\iota_{\partial_n})A_\pm\;,
\ee
where $A_\pm$ is any even/odd polyform. Abusing of the notation, sometimes we write expressions like $\buildrel\to\over\gamma\!{}^{m}A_\pm$ and $A_\pm\!\!\buildrel\leftarrow\over\gamma\!{}^{m}$, to be read as the Clifford map counter-image of (\ref{eq:gammaSlashA}).

A basis for the decomposition of $\wedge^{\bullet}T^*$ under the SU(3)$\times$SU(3) subgroup of O(6,6) defined by the `lowest weight states' $\Phi^0_\pm$ can be built acting with creators \cite{GMPT2, GMPT3, GenHodgeDec}:
\begin{equation}\label{eq:BasisDiamond}
  \begin{array}{c}\vspace{.1cm}
\Phi^0_+ \\ \vspace{.1cm}
\Phi^0_+\gamma^{i_2}  \hspace{1cm} \gamma^{\bar \imath_1}\Phi^0_+ \\ 
\Phi^0_-\gamma^{\bar \imath_2} \hspace{1cm} \gamma^{\bar \imath_1} \Phi^0_+\gamma^{i_2} 
\hspace{1cm} \gamma^{i_1} \bar \Phi^0_-\\
\Phi^0_- \hspace{1.2cm}\gamma^{\bar \imath_1}\Phi^0_-\gamma^{\bar \imath_2} 
\hspace{1cm} \gamma^{i_1}\bar\Phi^0_-\gamma^{i_2} 
\hspace{1.2cm}\bar\Phi^0_-\\
 \gamma^{\bar \imath_1} \Phi^0_-\hspace{1cm} \gamma^{i_1} \bar \Phi^0_+ \gamma^{\bar \imath_2} 
\hspace{1cm}\bar\Phi^0_-\gamma^{i_2}\\
\gamma^{i_1} \bar\Phi^0_+ \hspace{1cm}  \bar\Phi^0_+\gamma^{\bar \imath_2}\\
 \bar\Phi^0_+\\
  \end{array}\ 
\end{equation}
Each element of this `generalized diamond' transforms in a definite representation $(\bf{r},\bf{s})$ of \sst. We call $U_{\bf{r},\bf{s}}$ each of these subbundles of $\wedge^\bullet T^*$.

A basis for the decomposition under the SU(3)$\times$SU(3) structure defined by the $b$-transformed pure spinors $\Phi_\pm =e^{-b}\Phi^0_\pm$ is obtained simply acting with $e^{-b}$ on the above diamond (in the polyform picture).

One of the nice properties of the basis (\ref{eq:BasisDiamond}) is the orthogonality of its elements in the Mukai pairing: the only non-zero pairings are between elements in conjugate representations $(\bf{r},\bf{s})$ and $(\bf{\bar r},\bf{\bar s})$ of \sst.

The action of the operator $*\lambda$ can be easily evaluated using the Clifford map and eq.$\:$(\ref{eq:chiralgammaOngammas}):
\be\label{eq:*lambda}
\begin{picture}(10,11)(-15,1)
\put(0,0){\line(3,1){30}}
\end{picture}*\lambda(A) =-i\gamma \slash \!\!\!\!A\;.
\ee
Thus the result of the action of $*\lambda$ on each element of the diamond (\ref{eq:BasisDiamond}) is just $+ i$ or $-i$.

The Mukai pairing (\ref{eq:DefMukai}) between two forms can instead be evaluated via:
\be\label{eq:MukaiUnderClifford}
\langle A_k , C_{6-k} \rangle \;=\; \frac{i}{8}\tr(\gamma \:\slash\!\!\!\! A_k ^{\;T}\!\slas{\;\;C_{6-k}})vol_6\;.
\ee
For instance, for pure spinors $\Phi_\pm=e^{-b}\Phi^0_\pm$, with $\Phi^0_\pm$ built as bispinors, one finds
\be\label{eq:PSnorm}
i\langle \Phi_\pm,\bar\Phi_\pm\rangle \equiv i\langle \Phi^0_\pm,\bar\Phi^0_\pm\rangle = 8||\eta^1_\pm||^2||\eta^2_\pm||^2vol_6\,.
\ee

\section{Type IIA action with fluxes}\label{StandardIIAwithFluxes}

\setcounter{equation}{0}

In this appendix we make explicit the compatibility of the system of democratic EoM/Bianchi identities (with no localized sources) considered in section \ref{DemocraticFormulation} with the standard formulation of the type IIA action.\footnote{The problem of writing a supergravity action in the presence of general D-branes is studied e.g. in \cite{BelovMoore, deAlwisTransitions}. These papers also discuss a possible background independent formulation.} In doing so, we reconsider an issue already discussed in the \hbox{literature} \cite{KachruKashaniPoor, IIAModuliStabilization} concerning the expression for the Chern-Simons piece of the action when NS and RR background fluxes are switched on. We derive a general form of this Chern-Simons term by requiring consistency with the equations of motion.

In order to make contact with the standard formulation of (massive) type IIA supergravity, we need to break the democracy among the RR fields stated in section \ref{DemocraticFormulation}. Eliminating via the self-duality relations (\ref{eq:10dSelfDualityF}) the forms\footnote{In this appendix all the forms are ten dimensional. Since there is no risk of confusion, we omit the hat symbol over them.} $F_6, F_8, F_{10}$ from eqs.$\:$(\ref{eq:10dDemocraticRReom/Bianchi}) and (\ref{eq:EoMfor10dimB}), we are left with the following set of independent equations in terms of $H, F_0, F_2$ and $F_4$ only:
\be\label{eq:BianchiH}
dH=0\;,
\ee
\be
\label{eq:BianchiStandardIIA} d F_0=0  \qquad,\qquad d F_2 - H F_0 = 0 \qquad,\qquad  d F_4 - H\wedge F_2 =0\;,
\ee
\be\label{eq:EoMBstandardIIA}
d( e^{-2\phi} * H) - F_0 \wedge * F_2 - F_2\wedge * F_4 -\frac{1}{2} F_4\wedge F_4  = 0\;,
\ee
\be 
\label{eq:EoMStandardIIA} d *F_2 + H\wedge * F_4 = 0\qquad ,\qquad d * F_4 + H\wedge F_4 = 0\;.
\ee
In a topologically trivial background (where no fluxes can be switched on), the Bianchi identities (\ref{eq:BianchiH}) and (\ref{eq:BianchiStandardIIA}) are solved in terms of globally defined NS 2--form $B$ and 1-- and 3--form RR potentials $C_1$ and $C_3\,$:
\be
\label{eq:FieldStrWithoutFlux}  H=dB\quad,\quad F_0 = \mathrm{const} \quad,\quad F_2 =  dC_1 + B F_0 \quad,\quad F_4 = d C_3 - H \wedge C_1 + \frac{1}{2} B^2 F_0\;.
\ee
Now we can immediately check that the remaining equations (\ref{eq:EoMBstandardIIA}) and (\ref{eq:EoMStandardIIA}) correspond to the EoM for the potentials $B, C_1$ and $C_3$ descending from the standard massive type IIA (bosonic) action $S_{\mathrm{IIA}}$, with mass parameter $F_0$. Denoting $S_{\mathrm{IIA}}= S_{\mathrm{kinetic}}+ S_{\mathrm{CS}}$, we have (see e.g.$\:$\cite{Democratics}):
\be\label{eq:S_kinetic}
S_{\mathrm{kinetic}} = \;\frac{1}{2}\int \Big[e^{-2 \phi}\big( R*1 + 4 d \phi\wedge *d \phi - \frac{1}{2} H\wedge * H\big) -\frac{1}{2} \big(F_0\wedge * F_0 +  F_2\wedge * F_2  + F_4\wedge * F_4 \big) \Big],
\ee
\be\label{eq:S_CS}
S_{\mathrm{CS}} = -\frac{1}{4}\int \big[ BdC_3dC_3 + \frac{1}{3}F_0B^3dC_3 + \frac{1}{20}F_0^2B^5 \big]\;
\ee
(the $\wedge$ symbol is understood in $S_{\mathrm{CS}}$). Notice that the $F_0 =0$ limit yields the standard massless type IIA action \cite{PolchinskiVol2}.

Things become more subtle if one looks for general global solutions of the Bianchi identities (\ref{eq:BianchiH}) and (\ref{eq:BianchiStandardIIA}) on topologically non-trivial backgrounds, allowing for fluxes of the NS and RR field-strengths. In this case the expressions in (\ref{eq:FieldStrWithoutFlux}) are modified as follows ($F_0$ is still a constant parameter):
\bea
\nnb H &=& H^{\mathrm{fl}} + dB\;,\\
\nnb F_2 &=&  dC_1 + F_2^{\mathrm{fl}} + B F_0\;,\\
\label{eq:FieldStrWithFluxes} F_4 &=& d C_3 - H \wedge C_1 + F_4^{\mathrm{fl}}+ B\wedge F_2^{\mathrm{fl}}  + \frac{1}{2} B^2 F_0\;,
\eea
where the forms labeled with `${\mathrm{fl}}$' are defined as the non-exact parts of the solutions, satisfying the conditions
\be
H^{\mathrm{fl}}F_0 = 0 \qquad, \qquad dH^{\mathrm{fl}} =0 \qquad,\qquad d F_2^{\mathrm{fl}}=0 \qquad,\qquad d F_4^{\mathrm{fl}} - H^{\mathrm{fl}}\wedge F^{\mathrm{fl}}_{2} = 0\;.
\ee 
The first condition holds because if $F_0\neq 0$, then the Bianchi identity $dF_2 - HF_0 = 0$ implies that $H$ is exact and therefore $H^{\mathrm{fl}}=0$. In the expression (\ref{eq:modifiedCS}) below we will however keep both $H^{\mathrm{fl}}$ and $F_0$, also because the $F_0H^{\mathrm{fl}}=0$ constraint can be invalidated by the possible introduction of localized sources such as O6 planes,\footnote{In this case of course the action needs to be completed with the terms describing the couplings to the localized sources.} which modify the Bianchi identity for $F_2$ (see for instance \cite{IIAModuliStabilization, AcharyaBeniniValandro, VilladZwirner, deAlwisTransitions}).

We should now consider how the new expressions (\ref{eq:FieldStrWithFluxes}) for the field-strengths enter in the type IIA action. While we can simply substitute such new expressions into the kinetic terms (\ref{eq:S_kinetic}), the determination of the Chern-Simons action (\ref{eq:S_CS}) is more delicate. In \cite{KachruKashaniPoor} a modified form of the Chern-Simons term was obtained by requiring consistency with the structure of the expected 4d $N=2$ gauged supergravity after compactification on a Calabi-Yau three-fold, while in Appendix A of \cite{IIAModuliStabilization} it was deduced by properly modifying the M-theory Chern-Simons term in order to accomodate for a 4-form flux, and then performing the reduction to ten dimensions. 

Here we propose a general expression for $S_{\mathrm{CS}}$ by imposing that the equations of motion derived from the action still have the form (\ref{eq:EoMBstandardIIA}), (\ref{eq:EoMStandardIIA}). We can see that this requirement is satisfied if we preserve the form (\ref{eq:S_kinetic}) for $S_{\mathrm{kinetic}}$, and modify the Chern-Simons term as follows:
\bea
\nnb S_{\mathrm{CS}} &=& -\frac{1}{4}\int \Big[ C_3H^{\mathrm{fl}}(dC_3+2F_4^{\mathrm{fl}}) + B(dC_3 + F_4^{\mathrm{fl}})(dC_3 + F_4^{\mathrm{fl}}) +  B^2F_2^{\mathrm{fl}}(dC_3 + F_4^{\mathrm{fl}})\\
\label{eq:modifiedCS} && \;\qquad + \;\frac{1}{3}B^3F_2^{\mathrm{fl}}F_2^{\mathrm{fl}} + \frac{1}{3}F_0B^3(dC_3 + F_4^{\mathrm{fl}}) + \frac{1}{4}F_0B^4F_2^{\mathrm{fl}} + \frac{1}{20}F_0^2B^5 \Big]\;.
\eea
This expression not only is in agreement with the ones given in \cite{KachruKashaniPoor, IIAModuliStabilization}, but also extends it to the case of non-vanishing $F_2^{\mathrm{fl}}$, which was not considered in those papers.

One can lastly verify that the field-strengths $H, F_2, F_4$, as well as the complete action $S_{\mathrm{IIA}}$, are invariant under the following globally defined gauge transformations involving the $k$--form (infinitesimal) parameters $\Lambda_k$: 
\be
\delta B\, =\, d\Lambda_1\quad,\quad
\delta C_1\, =\, d\Lambda_0 - \Lambda_1 F_0 \quad,\quad
\delta C_3\, =\, d\Lambda_2 - H\Lambda_0 - \Lambda_1(F_2^{\mathrm{fl}} + BF_0)\;.
\ee
The EoM (\ref{eq:EoMBstandardIIA}), (\ref{eq:EoMStandardIIA}) are of course gauge-invariant due to the invariance of the field-strengths.


\end{document}